# Atomic-Precision Fabrication of Quasi-Full-Space Grain Boundaries in Two-Dimensional Hexagonal Boron Nitride

*Xibiao Ren, Xiaowei Wang, Chuanhong Jin\**

State Key Laboratory of Silicon Materials, School of Materials Science and Engineering Zhejiang University, Hangzhou, Zhejiang 310027, China.

ABSTRACT

Precise control and in-depth understanding of the interfaces is crucial for the functionality-oriented material design with desired properties. Herein, via modifying the long-standing bicrystal strategy, we proposed a novel nanowelding approach to build up interfaces between two-dimensional (2D) materials with atomic precision. This method enabled us, for the first time, to experimentally achieve the quasi-full-parameter-space grain boundaries (GBs) in 2D hexagonal boron nitride (h-BN). It further helps us unravel the long-term controversy and confusion on the registry of GBs in h-BN, including i) discriminate the relative contribution of the strain and chemical energy on the registry of GBs; ii) identify a new dislocation core- Frank partial dislocation and four new anti-phase boundaries; and iii) confirm the universal GB faceting. Our work provides a new paradigm to the exploiting of structural-property correlation of interfaces in 2D materials.

Grain boundaries (GBs), an inevitable structural imperfection in polycrystalline two-dimensional (2D) materials, behave as one-dimensional line defects that connect differently oriented grains of the same material[1,2]. Extensive studies have confirmed that the inherent structures of GBs (e.g., misorientation angles and interior atomic arrangements) determine the mechanical, thermal, and electrical properties of 2D materials[2–5]. Thus, designing and tailoring GBs at atomic scale is one of the utmost goals for advancing fundamental research and industrial applications of 2D materials. Interface generation, i.e., the formation of GBs in 2D materials is hard to control during the growth[6] (e.g., the widely adopted chemical vapor deposition (CVD))[6], due to the stochastic nucleation of individual grain at high temperatures. A variety of approaches have been developed towards the controlled GB formation that includes the symmetry design of the growth substrates[7], control of the grain sizes[8], and deployment of thermal excitation and beam irradiation[9], etc. Despite all those efforts, GB formation is still hard to control precisely in CVD due to the complex thermodynamics and kinetics. For instance, hydrogen passivated growing edges leads to the formation of overlapping GBs rather than covalently bonded GBs in CVD grown 2D hexagonal boron nitride (h-BN)[10]. More importantly, GBs with desired misorientation angles (requiring certain precision, e.g., <0.5°) are hard to realize by those approaches. For example, only several specific misorientation angles can be achieved via controlling the symmetry of substrate, e.g., 30° for GBs in graphene; And the probability is too low to find a GB with specific misorientation angle in a polycrystalline CVD film. Thus, up to date, the atomic-precision design and fabrication of GBs in 2D materials remain a challenge.

Bicrystal technique has so far served as the most accurate way to prepare the GBs for decades[11], in which two separate single crystals with desirable relative crystallography orientation are joined together to form a planar interface (e.g., GB), as sketched in Figure 1a. The success of this method repeated over a wide range of materials including metals[12], ceramics[11], but not in 2D materials yet (Figure 1b). It might be attributed to two technical barriers: i) the flexibility of 2D membrane makes the manipulation hard to operate, and ii) lacking precise crystallographic information of the grains, i.e., the misorientation angle, relative sliding, edge structure and cleanness prior to the join, where structural characterization and interfaces fabrication at atomic scale are necessary.

**RESULTS AND DISCUSSION**

**Quasi-bicrystal nanowelding in 2D materials**

In this work, by modifying the long-standing bicrystal technique, we proposed a novel quasi-bicrystal nanowelding approach that enables the design and fabrication of GBs in 2D materials with atomic precision. The whole process was illustrated in Figure 1c. First, a local, bilayer region with an interlayer twisting angle-$\theta_t$ on 2D material was chosen. Via the electron beam induced sputtering inside the transmission electron microscope (TEM)[13], we then created two adjacent holes: one in top layer and the other in bottom layer separately, during which the location and the size of the holes could be controlled accurately down to atomic level as enabled by the employed spherical aberration-corrected electron optics; see Figure 1c (i) to (ii). As the holes expanded in size upon prolonged electron beam irradiation, their edge lines approach each other though sit in different layers (top and bottom layers, respectively). And finally they joined spontaneously and seamlessly once they met, leading to the formation of a GB in between; see Figure 1c (iii) to (iv).

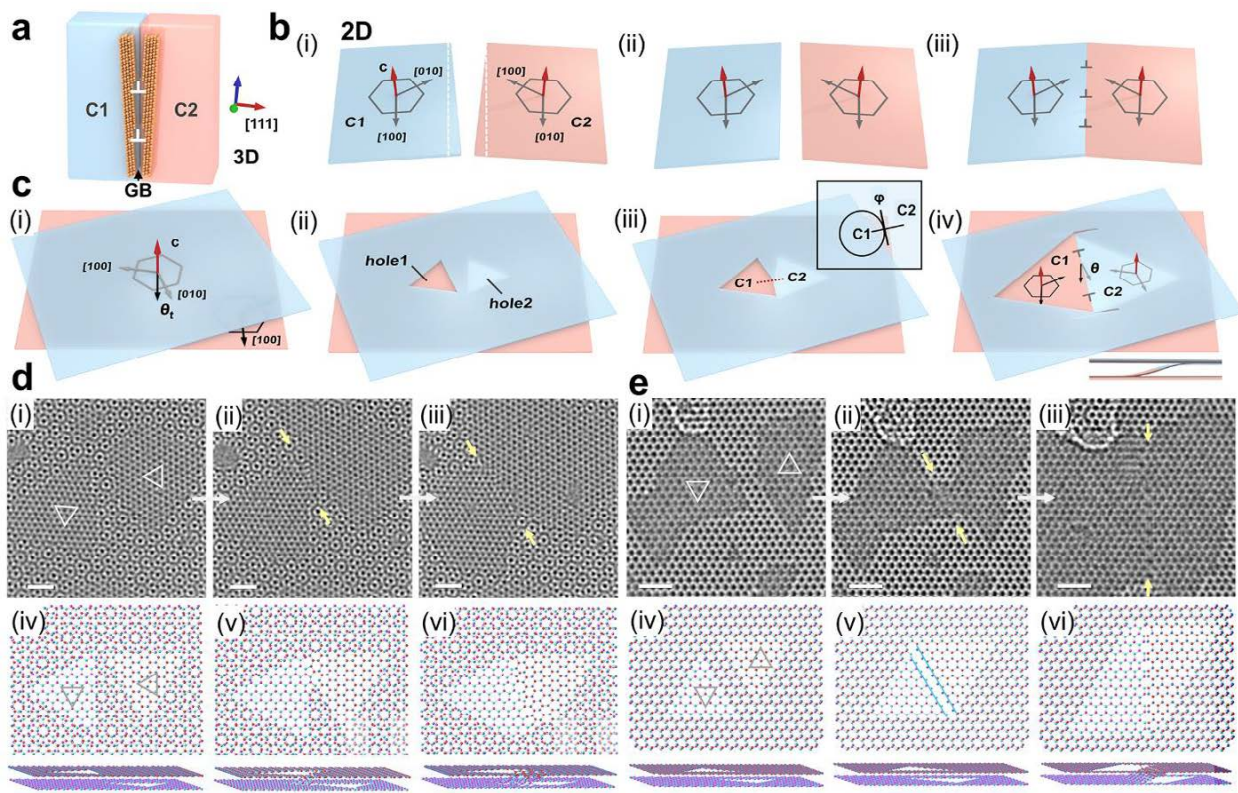

Figure 1. Interfaces fabrication. (a, b) Schematic diagram of the traditional bicrystal method in bulk materials and hypothesized illustration of the bicrystal method in 2D materials. "C1" and "C2" represent crystal 1 and crystal 2, respectively. (c) Schematic illustration of the quasi-bicrystal method proposed in this study. (d, e) Sequential HRTEM images showing the GB fabrication, (d) for a $|\theta|=30°$ GB and (e) for a $|\theta|=60°$ GB, and their corresponding atomic models are shown in below (top: planar view; bottom: cross-sectional view), respectively. The as-formed GBs were marked by yellow arrows. The red (purple) and blue (cyan) balls represent boron (B) and nitrogen (N) atoms in the top layer (bottom layer), respectively. The triangles in d (i), (iv) and e (i),(iv), connecting the nitrogen sub-lattices, represent the grain orientations. Scale bars: 1 nm.

We then experimentally demonstrated the applicability of the proposed method in preparing homo-junctions and GBs in 2D materials, taking 2D hexagonal boron nitride (h-BN) as an example. Owing to its binary-hexagonal-lattice nature, 2D h-BN features a richer variety of GB configurations compared with that in graphene[14]. CVD grown h-BN flakes containing overlapping (bilayer) regions with different interlayer twisting angles ($0°≤θ_t≤120°$) were chosen for *in-situ* nanowelding. Sequential HRTEM images shown in Figure 1d (i to iii) illustrates the formation dynamics of a 30°-GB. Under the controlled electron beam irradiation and sample temperature of 1073 K, the hole size and edge length increased gradually via the atom-by-atom sputtering. Once the two nearby edges (with a ~1 nm length) met each other (remember again they are located in different layers), a GB formed spontaneously, which connects the top and bottom layers seamlessly; see Figure 1d (ii). Another example showing the formation of a 60°-GB was displayed in Figure 1e, and details were shown in Movie S1&S2.

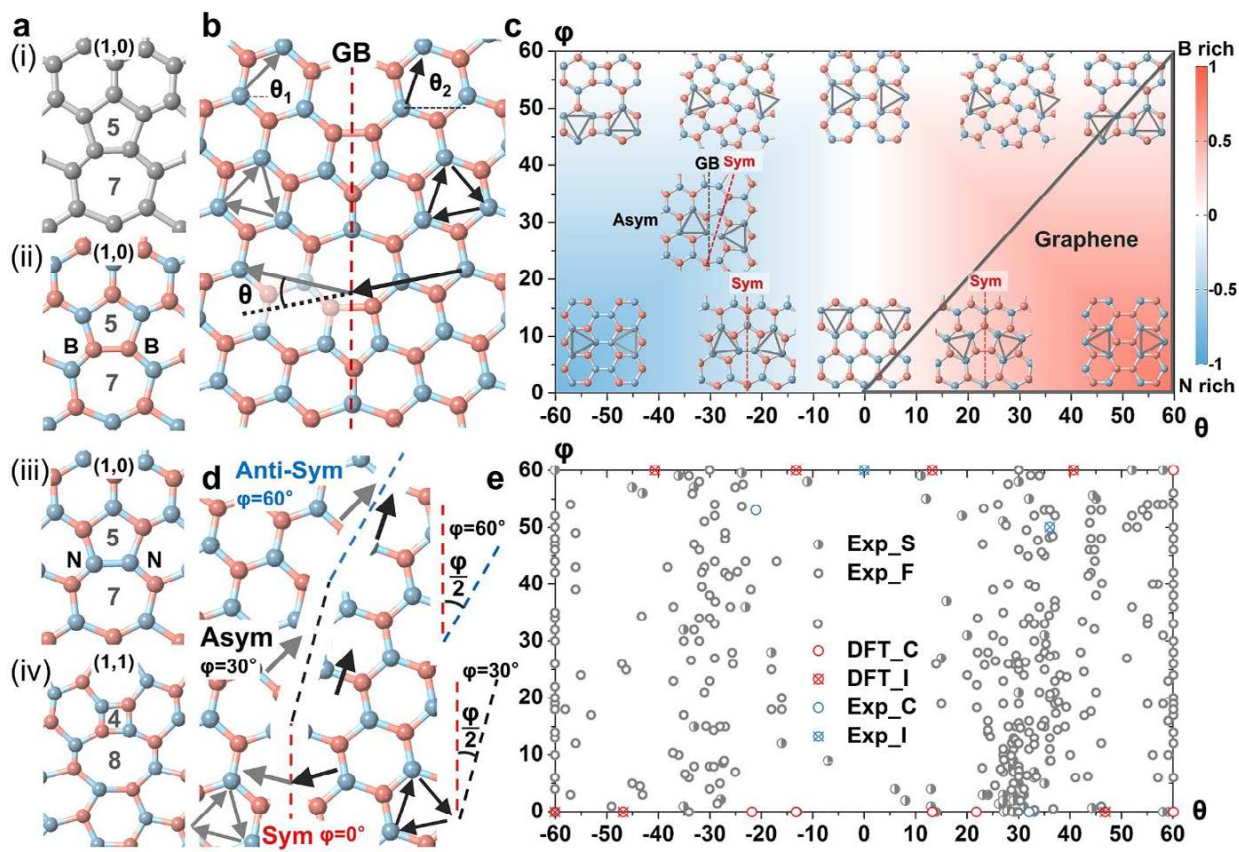

Figure 2. The parameter space of GBs in 2D binary hexagonal materials. (a) Atomic structural models showing the 5|7s and 4|8 dislocations in graphene and binary hexagonal lattice. Grey: carbon, Red: boron, blue: nitrogen. Refer to Figure S7 for more details. (b) Definitions of $θ_1$, $θ_2$, $θ$, respectively. (c) The parameter space of GBs in 2D h-BN, with its depth of color represents the degree of enrichment of the B

(red) or N (blue) element in the GB. Thus, the left side represents N-rich GBs and the right side is for B-rich GBs. The grey triangle area indicates the space of GBs in graphene, which is 1/4 that of 2D h-BN. Several typical GB structures are displayed. The red dash lines represent symmetric GBs, and grey triangles in structural models indicate the orientations of two grains. (d) The relation between φ and the inclination angle of the facet line to the sym-facet (φ=0°). Three typical facets are displayed: φ=0° (sym, red dash line, same as that in b), φ=30° (asym, black dash line) and φ=60° (anti-sym, blue dash line). (e) Experimentally obtained 437 GB-(θ, φ)s and GBs from previous studies. "Exp_S" and "Exp_F" represent straight and faceted GBs obtained in this work. "DFT_C/ I" and "Exp_C/ I" represent GBs predicted by theory[14] and observed in previous experiments[19–21], where "C" and "I" represent consistent and inconsistent with our work, respectively.

To fully describe the GBs in 2D materials, two macroscopic parameters-θ and φ are required. θ is the misorientation angle of the two subject domains as shown in Figure 2b, and φ is two times the inclination angle of the boundary line to the symmetric configuration (φ=0°) as illustrated in Figure 2d. The derived parameter-space is displayed in Figure 2c. In our quasi-bicrystal approach, θ is pre-determined by the interlayer twisting angle between the top and bottom layers-$θ_t$ (Figure 1c (i), (iv)), which can be controlled with an accuracy of 0.1° to 0.5° depending on the small variation of grain rotations. The other parameter-φ of a GB usually follows the inclination of the as-formed GB as illustrated by Figure 1d (ii), (iii). Hence, φ can also be controlled to some extent by controlling the relative positions of the two holes, which can be then controlled by e-beam irradiating areas; see inset in Figure 1c (iii). Depending on the extent of structural relaxation of a GB just after welding, which is further related to θ and the length of GB, the control accuracy of ~6° to ~30° can be reached for most GBs in practice. More details on the accuracy and the control of φ were discussed in Method Section. It is worthwhile to mention that some GBs may be metastable upon joining, and thus undergo further structural relaxation until reaching an energy-favored configuration, such as the case for the 60°-GB shown in Figure 1e (ii), GB rotated itself for 30° clockwise (to Figure 1e (iii)). All in all, θ and φ in our quasi-bicrystal nano-welding of 2D materials are crystallographically equivalent to those used in the traditional bicrystal technique for bulk materials.

Our ~250 joining attempts for making h-BN/h-BN homo-junction all succeeded under our typical experimental conditions, and the attained GB length ranged 1-26 nm, suggesting the high reproducibility and feasibility of the quasi-bicrystal nanowelding approach. Nevertheless, the obtained GBs were demonstrated to be equivalent to regular planar GBs provided they are long enough, e.g., longer than 5.5 nm as explained in detail in Figure S1, S2. Bonding between the under-coordinated edge atoms along the hole edges occurs spontaneously as a result of the reduction in the total energy via minimizing the number of dangling bonds, but at an expense of energy gain due to the introduction of a GB into the lattice of 2D h-BN. Note that, a high sample temperature is necessary to retain the edge cleanness, and depress the edge passivation[10], and thus promote the GB formation.

Our method also works well on fabricating hetero-junctions between different 2D materials such as graphene and h-BN, as demonstrated in Figure S3. The atomic resolution TEM and electron energy loss (EEL) spectra confirmed the successfully building up a seamless graphene/h-BN interface. In principle, this method should also be extendable to other 2D materials like transition metal dichalcogenides (TMDs). Deployment of mechanical exfoliation plus dry transfer can further ensure the accurate control over the constitute 2D materials such as their relative interlayer twisting[15,16].

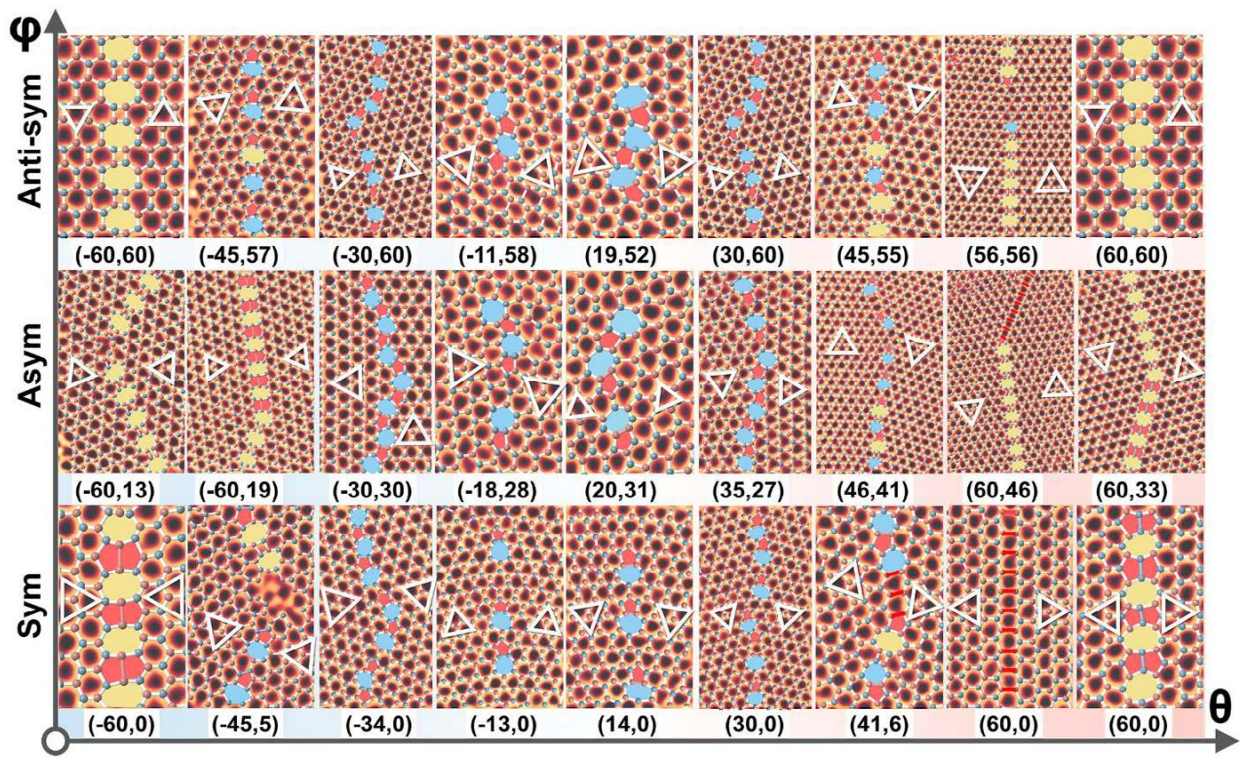

Figure 3. The experimentally achieved quasi-full-parameter space GBs in 2D h-BN. Atomic HRTEM images of the as-formed GBs covered with their corresponding structural models (Experimental images alone were displayed in Fig. S9). Pentagons, heptagons, and octagons are filled with red, blue and orange, respectively. The white triangles indicate the domain orientations. A GB with $\theta$ and $\varphi$ was labeled as GB-$(\theta,\varphi)$, e.g., GB with $\theta=14°$ and $\varphi=0°$ was labeled as GB-(14,0). Note that GB-$(\theta,60)$ are identical to GB-$(-\theta,60)$ under space definition. More details are shown in Figure S10-11.

**Quasi-full-parameter space of GBs in 2D h-BN**

The quasi-bicrystal method proposed in this study, for the first time, unlocked the door to experimentally achieve a quasi-full-parameter space GBs (~430 GBs) in 2D h-BN, as displayed in Figure 2e and Figure 3, where a GB with θ and φ was labeled as GB-(θ,φ) for brevity. Due to its binary nature, the macroscopic parameter space of GBs in 2D h-BN is four times in size of that for graphene, and thus bring in more complexities. The formation energy ($E_f$) of a GB in h-BN can be estimated as $E_f=E_s+E_c$[17,18], where $E_s$ is strain energy and $E_c$ is chemical energy owing to the presence of homoelemental bondings (B-B, N-N bonds) in a GB. As a result, the competition between strain and chemical energy leads to more complex θ- and φ- dependent GB configurations in 2D h-BN, in comparison to that for graphene. This complexity was still poorly understood mainly due to the limited experimental results[19–21] (indicated by Exp_C/I in Figure 2e) and was underestimated theoretically (indicated by red solid circles and marked as DFT_C and DFT_I in Figure 2e. "C" and "I" represents that the theoretical results were consistent and inconsistent with our experimental results, respectively). For example, even the dislocation cores that constitute a GB are still under debate with controversial results shown among different experiments[19–21].

The unprecedented quasi-full-parameter-space GBs enabled us to discriminate the relative contribution of strain energy ($E_s$) and chemical energy ($E_c$) to the registry of GBs. Due to the limited scope of this manuscript, hereafter we will focus on three primary experimental findings: i) θ and φ dependent dislocation cores; ii) anti-phase boundaries (APBs), and iii) GB faceting. For the sake of discussions and taking the small variation of φ in experiments into consideration, GBs can be divided into three classes, that is, symmetric GBs (sym-GBs) with φ<6°, asymmetric GBs (asym-GBs) with 6°≤φ≤54° and anti-symmetric GBs (anti-sym GBs) with 54°<φ<60°.

As summarized in Figure 3 and Figure S8, strain energy governs the GB structures for |θ|<~38°, similar to that of graphene (Figure S11), while chemical energy prevails for |θ|>~38° GBs and that leading to diverse dislocations cores in high-angle GBs. According to the previous theoretical calculations[14], two types of dislocation cores were predicted in 2D h-BN: i) the 5|7 dislocation with a Burger's vector $\vec{b}_{(1,0)}$ ($|\vec{b}_{(1,0)}|$=2.50 Å) as shown in Figure 2a (ii, iii); and ii) the square-octagon (4|8) dislocation with a Burger's vector $\vec{b}_{(1,1)}$ ($|\vec{b}_{(1,1)}|$=4.33 Å) (Figure 2a (iv)). Compared to 4|8, 5|7 possesses smaller strain energies ($E_s$) as $E_s$ is proportional to $|\vec{b}|^2$ ($E_s \propto |\vec{b}|^2$), but also higher chemical energies ($E_c$) owing to the presence of energy unfavorable homoelemental bonding, i.e., B-B or N-N bonding. It was theoretically predicted that 5|7s constitute sym-GBs while 4|8s form anti-sym GBs, respectively[14]. However, in contrary to those

theoretical predictions, we could derive following rules for GB configurations from our experimental results: i) for $|\theta|<\sim38°$, both sym- and anti-sym GBs are formed only by 5|7s; ii) for $\sim38°<|\theta|<47°$, GBs are comprised of 5|7s and a new type of dislocation core-Frank partial dislocations (5|84|7s or 5|84|...|84|7s) as shown in Figure 4c, d, which was not reported previously. Note that $\theta>38°$ sym-GBs can also consist of only 5|7s, as exampled by the GB-(41,5) in Figure S9; iii) for $\sim47°<|\theta|<60°$, anti-sym GBs only consists of Frank partial dislocations, while sym-GBs and asym-GBs contain both 5|7s and Frank partial dislocations; and iv) no 4|8 dislocation core was observed (served as $\vec{b}_{(1,1)}$ dislocations).

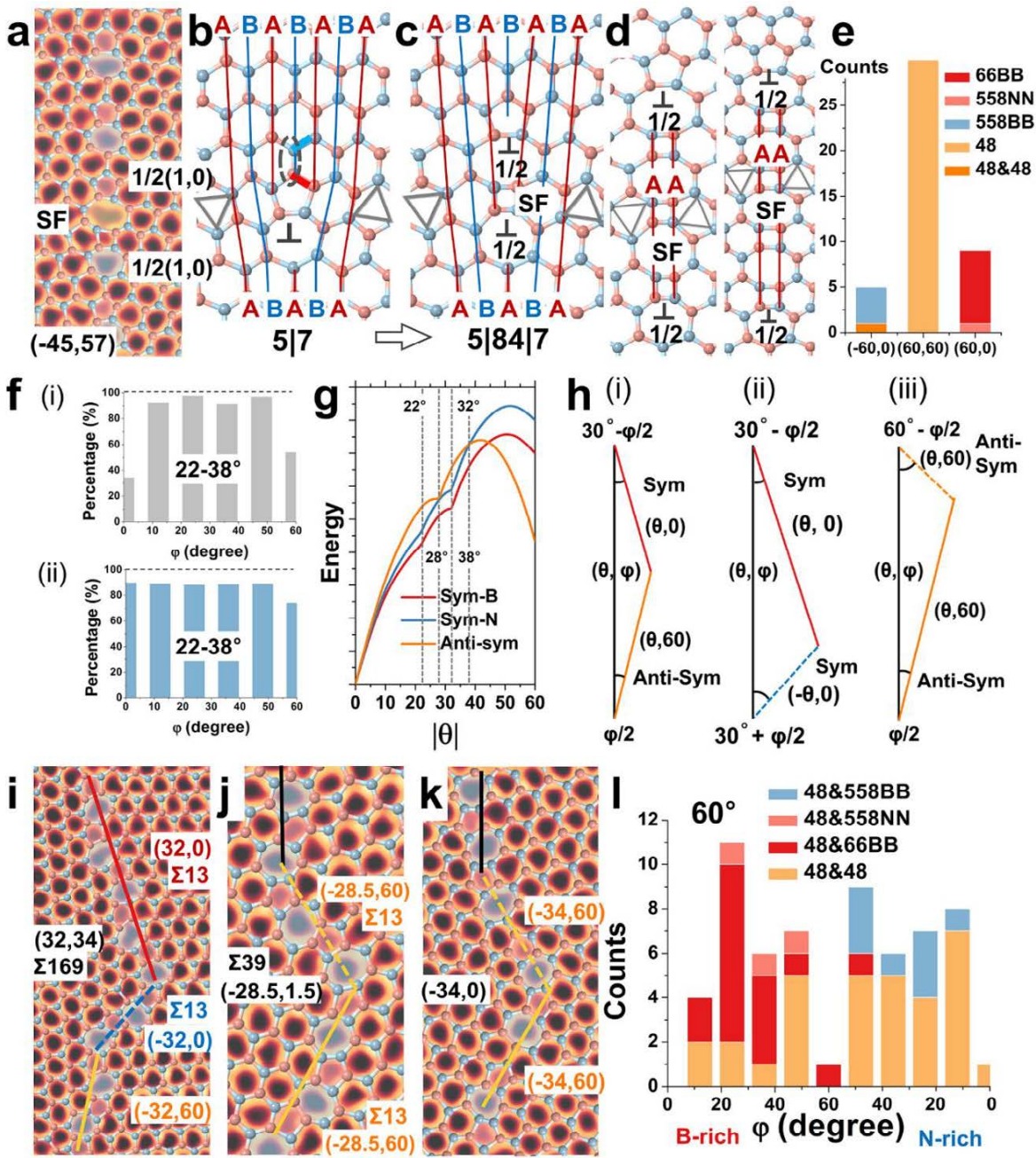

Figure 4. Frank partial dislocations and GB faceting in 2D h-BN. (a-c) Experimental Frank partial dislocations and structure models showing the transformation of perfect dislocation (5|7) to a Frank partial dislocation pair 5|84|7 and stacking fault. "A" (red) and "B" (blue) represent (110) lattice planes. The "SF" in (C) stands for stacking fault, and "1/2" represents 1/2(1,0) dislocation. (d) Structural models of other partial dislocation pairs with longer stacking faults. (e) Statistics of APBs. "48&48" represents two 48 facets. (f) Statistics of GB faceting. (i-ii) The proportion of faceted GBs to all the GBs, the average proportion of the length of symmetric/anti-symmetric facets to the overall length of a GB with a φ range (e.g, 6°<φ<18°), respectively. (g) Formation energy of symmetric and anti-sym facets. Refer to Figure S13, 15 for more details. (h) Schematic diagram for three types of faceting with the shortest total length of GBs among numerous combinations of symmetric/anti-symmetric facets. (i-k) Experimental results showing GB faceting in three representative GBs: asym-GB, CSL sym-GB (nearly), non-CSL sym-GB. (l) Statistics of faceting types in APBs. For example, "48&558BB" represent one facet-48 and one facet-558BB.

Frank partial dislocations well solve the dilemma of competition between chemical and strain energy for $|\theta|>38°$ GBs, and are energetically favorable compared with 5|7s and 4|8s in anti-sym GBs for large $|\theta|$. The Frank partial dislocations in Figure 4c are formed via dislocation dissociation from a perfect (1,0) dislocation as (1,0)→1/2(1,0)+1/2(1,0), and this reaction was accomplished by the removal of a B-N pair (as shown by the dotted circle in Figure 4b). Meanwhile, a stacking fault constituting 4|8s forms between two Frank partials along with the removal of homoelemetal bonding (Figure 4b to 4c), whose stacking sequence is ABAABA..., where A and B represent (110) lattice planes. Note that 4|8s are not serving as the dislocation cores here. The total energy of two 1/2(1,0) Frank partial dislocations adds up to $1/2b_{(1,0)}^2$ as $E_{1/2(1,0)}+E_{1/2(1,0)}=(1/2b_{(1,0)})^2+(1/2b_{(1,0)})^2$, corresponding to half of the strain energy of a perfect (1,0) dislocation. Moreover, 4|8 stacking faults are energetically favorable compared to the hexagonal lattice containing B-B and N-N homoelemental bonding (Figure 4b)[17]. Hence, the dislocations dissociation and the associated introduction of stacking fault reduce the GB formation energy in 2D h-BN, which is distinctive from that in elemental materials[22]. It should be noted that the critical misorientation angle $|\theta|$-38° was derived based on two facts. First, according to the coincidence site lattice (CSL) theory[23] and considering the elimination of homoelemental bonding as discussed in Figure S8, the Frank partial dislocations can only be formed when $|\theta|>\sim 38°$; Second, contributions of chemical energy to the total GB energy becomes prevailing when $|\theta|>\sim 38°$ as shown in Fig. S13.

GBs with |θ|=60° also denoted as anti-phase boundaries (APBs), are unique in binary 2D materials due to the absence of inversion symmetry. Up to five stable APBs were theoretically predicted in 2D h-BN that are comprised of topologically trivial configurations, i.e., with a Burgers vector $|\vec{b}|=0$[17], but so far direct experimental evidence for these APB configurations remains rather limited. Herein, four of them were *in-situ* prepared and unambiguously identified as shown in Figure 3. Of them, the most common one is GB-4|8s (GB-(60,60), comprised of 4|8s), as evidenced by the statistics in Figure 4e. Detailed configurations of APBs depend on the crystallography orientation. For GBs lying along the armchair direction (φ=60°), they comprise of 4|8s. While GBs lying in zigzag direction (φ=0°) consist of 66-BBs (GB-(60,0), hexagonal lattice with B-B bonds, as shown in Figure S12), 558-NNs (GB-(60,0), with N-N bonds) and 558-BBs (GB-(-60,0), with B-B bonds), respectively. Given the crystal symmetry of 2D h-BN into consideration, 66-NNs (GB-(-60,0), hexagonal lattice with N-N bonding) are also expected but not found experimentally, even though the in-situ experiments were conducted in a relatively nitrogen-rich environment due to the preferable sputtering of boron atoms upon e-beam irradiation[24]. This finding allows us to conclude that N-N bonding is less stable than B-B bonding in 2D h-BN.

**GB faceting in 2D h-BN**

We then discuss the GB faceting in 2D h-BN, which is poorly understood in 2D materials with only a few reports related to the meandering phenomenon of GBs in graphene[25–27], and the universality remains unexplored. As shown by the black open circles in Figure 2e and exampled in Figure 4i-k, GBs usually contain different facets rather than being solely straight[28]. Statistical analysis further showed that more than 90% asym-GBs with 22°<|θ|<38° (Figure 4f (i)) and ~100% asym-GBs with 38°<|θ|<60° (Figure S14) are faceted. To our surprise, faceting also exists abundantly in high-symmetry GBs, i.e., 34% of the sym-GBs and 54% of the anti-sym GBs; see Figure 4f (i). Typically, a GB prefers to dissociate itself into shorter symmetric/anti-symmetric facets. Note that the definitions of symmetric/anti-symmetric facets in GBs are the same as that of GBs. As shown in Figure 4h and the statistics in Figure 4f (ii) and Figure S14, the length population of symmetric/anti-symmetric facets reaches >~80% for 22°<|θ|<47° GB, and ~100% for |θ|>47° GBs, respectively.

The origins of GB faceting in h-BN and other 2D materials can be well understood as the result of energy decrease due to the formation of energy-favorable facets prevailing the energy increase caused by the increasing of GB length. In short, it can be described as $\sum E_i L_i < E_0 L_0$, where $E_i$, $L_i$ are the energy per unit length of facet i and the length of facet i, respectively, i=1,2… represents different facets, in particular,

i=0 stands for the straight GB (as shown by the black lines in Figure 4h). So again, the competition between strain energy and chemical energy of a GB continues in GB faceting in 2D h-BN, especially for a large |θ| GB where the energy difference among B-rich, N-rich sym-facets, and anti-sym facets become prominent as shown in Figure 4g.

For GBs with |θ|<38°, GB faceting is mainly driven by strain. As described by CSL theory[23], the GB energy-E(θ,φ) usually has cusps at specific (θ,φ)s with high symmetry, corresponding to low Σ values, where Σ is the inverse density of coincidence sites (see Figure S8 for more details). For example, the asymmetrical GB-(32,34) shown in Figure 4i which is close to GB-(32.2,27.8) with a Σ=169 tends to break into two neighboring symmetric facets: GB-(32.2,0)s (red line, Σ13 GBs) and GB-(-32.2,0)s (dotted blue line, Σ13 GBs). Similarly, the symmetric GB-(28.5,1.5) with Σ=39 shown in Figure 4j broke into smaller Σ facets (two Σ13 facets). More generally, for sym-GBs with an arbitrary θ, i.e., the non-CSL GBs as shown in Figure 4k, where the dislocation cores are not well separated along the GB, the appearance of faceting therein can eliminate the local concentration of strain effectively (see the evidence from strain maps in Figure S16). Furthermore, the energy difference between B-B and N-N bonding also influence the detailed registry of GB faceting. For example, the GBs shown in Figure S18 preferred to undergo the faceting with a higher density of B-B bonds.

Chemical energy rules faceting types in |θ|>38° GBs in the following aspects: First, GBs dissociate preferably into anti-sym facets which consist of Frank partials for |θ|<60° or GB-48s for |θ|=60°, as exampled by GB-(46,41) and GB-(-60,13) in Figure 3; Second, there exist asymmetrical faceting of GBs with θ<0° and θ>0°, i.e., the most common facets combination is B-rich sym-facets&anti-sym facets when θ>0° while anti-sym facets&anti-sym facets when θ<0°, as evidenced by Figure 4l and Figure S19. The preferred anti-sym facets&anti-sym facets type in θ<0° GBs with longer GB total length (longer than that of anti-sym facets&N-rich facets) further prove the unstable N-N bonds. Interestingly, GB-(-60,0) can also be formed by two 48s facets (Figure 4e and its structure in Figure S9), similar to the GB-(-60,13) in Figure 3.

**Discussions and conclusions**

Considering the importance of chemical potential in 2D binary materials, we further discussed its effects on GB configurations on the following aspects. The chemical potential provides an opportunity to engineer GB structures in binary systems and even their total lengths for asym-GB and sym-GBs via changing the energetically favored facets. The chemical potential can be tuned during the synthesis as proposed by theoretical resuts[29]. However, due to the nonpolar nature of anti-sym GBs, the change of chemical potential has no effects on their structures. It should also be noted that the variation of chemical potential only leads to the asymmetric behaviors of GBs with θ>0° (right space in Figure 2c) and θ<0° (left space in Figure 2c). The main conclusions on the registry of GBs with different |θ|, e.g., Frank partial dislocations appear in high-angle GBs, will still maintain.

In summary, the quasi-bicrystal method opens a new area for the parameter-controlled topological design at atomic scale in 2D materials[30], and serves as a powerful approach towards the function-oriented design of interfaces. The binary 2D materials bring much more complexities in GBs that are beyond expectation. Understanding of relative contributions of strain and chemical energy in GBs which will greatly enrich our limited knowledge of GBs in 2D binary lattices. The demonstration of the universality of GB faceting in both numbers and diversities (asym-GBs and sym-GBs/anti-sym GBs, CSL, and non-CSL GBs) will renew the understanding of GBs in 2D polycrystalline materials and may offer new insights into structure-related GB properties such as mechanical strength, electronic transport and GB migration, etc. For instance, the GB faceting may lead to the transition from normal to abnormal grain growth[31].

**Experimental Methods**

Parameter-space. The parameter-space of GBs in 2D materials consists of two macroscopic degrees of freedom: θ and φ, and two microscopic degrees of freedoms: the relative sliding between two grains along x, y directions, respectively. Since the sliding can be effectively eliminated in the relaxed GBs obtained in our experiments, and thus their contribution will not be considered here. To distinguish two different atomic species (B and N) at the GB, vectors connecting one atomic species (nitrogen site here) in two grains in a clockwise direction are used (see Figure 2b in the main text). There exist pairs of $\theta_1$, $\theta_2$ that satisfy $(\theta_1-\theta_2)\in[-60°, 60°]$, where $\theta_1,\theta_2\in[0°, 360°]$. We defined $\theta=(-1)^m(\theta_1-\theta_2)\in[-60°, 60°]$ and $\varphi=\{(-1)^n(\theta_1+\theta_2)+120°l\pm240°k\}\in[0°,60°]$, where $m=[(\theta_1+\theta_2+60°)/120°]$, $n=[(\theta_1+\theta_2)/60°]$, and $k=0,1,2$, respectively. The notation [x] is to get the integer part of x (x is a real number), and l=0 if m is even while l=1 if m is odd. Our purpose to introduce the parameters-m, n, k, and l was to exclude the repetition of parameter space. The left part (N-rich) and the right part (B-rich) of the parameter space mirror each other if ignore the elemental species, which makes it clear to compare the B-rich and N-rich GBs, as shown in Figure 2c. Note that, (θ,60)-GBs are identical to (-θ,60)-GBs under our definition. For example, a (60,60)-GB is the same as a (-60,60)-GB. The parameter space can be applied to 2D h-BN and other binary hexagonal lattices such as monolayer transition metal dichalcogenides (TMDs).

We defined another parameter $p_{Bi}=(-1)^{i+j}(1.5-(\theta_i\pm120°k)/60°)$ to describe the polarity of boron element of the grain, where i=1, 2 represent the left and right grain, $j=[(\theta_i+30°)/60°]$, k=0,1,2… to keep $(\theta_i\pm120k)$ $\in(30°,150°)$, respectively. The notation [x] is to get the integer part of x (x is a real number). Accordingly, the polarity of nitrogen elements was defined as $p_{Ni}=1-p_{Bi}$. Hence, the polarity of a GB was defined as $P=((p_{B1}+p_{B2})-(p_{N1}+p_{N2}))/((p_{B1}+p_{B2})+(p_{N1}+p_{N2}))=p_{B1}+p_{B2}-1$, which was shown by the color bar in Figure 2c in main text. As a special case, P=0 represents the nonpolar condition such as anti-sym GBs (φ=60°).

Theoretically, GBs can be divided into three classes: i) symmetric GBs when φ=0°; ii) asymmetric GBs when 0°<φ<60°; iii) anti-sym GBs when φ=60°. Due to the fabrication and measurement errors, variations of φ should be considered. The inclination of sym-GB to asym-GB by φ will result in a shift of Ltan(φ/2) between the two ends of a GB, where L is the GB length. Suppose L= 5 nm and φ= 6°, it gives a 0.26 nm shift, almost the same as the lattice constant of h-BN (2.5 Å). Such a shift was very likely to occur in our experiments, causing a small variation in φ. Therefore, we approximately treated GBs with φ<6° (φ>54°) as sym-GBs (anti-sym GBs).

**Preparation of h-BN samples and h-BN/graphene heterostructures**. h-BN flakes were prepared by a low-pressure CVD, slightly modified from the previous work[32]. Vertical h-BN/graphene heterostructures were prepared via the standard wet transfer process of the CVD products, during which h-BN monolayers were loaded onto another pre-transferred monolayer graphene flake.

**TEM sample preparation and characterization**. Following the PMMA assisted electrochemical bubbling methods, as grown h-BN and graphene samples were transferred onto a TEM heating chip (Wildfire Nano-Chips XT, DENSolutions). In situ HRTEM was carried out in a TEM (Titan $G^2$ 80-300, FEI) operated at 80 kV. This microscope was equipped with a spherical aberration corrector on the imaging side and a monochromator that narrows the energy spread to ~0.15 eV, and thus the attainable spatial resolution at low voltages was largely improved. HRTEM images were recorded using a positive spherical aberration coefficient (C3) of ~5 μm and a negative defocus close to the Scherzer condition, under which the lattice atoms appear dark in h-BN monolayers. The electron dose rate was about $1\times10^6$ e/nm$^2$s for imaging, and the exposure time of each frame was 1 second. HRTEM images were acquired by a BM-UltraScan CCD (Gatan). Electron energy loss (EEL) spectra were collected with a spectrometer (Quantum 963, Gatan).

Boron and nitrogen lattice atoms were discriminated by analyzing the crystallography of the triangular holes as generated by the electron irradiation, since these hole edges were assigned/determined to be nitrogen-terminated due to the preferable sputtering of boron atoms[24,33]. Most of the holes remain in a triangular shape as the sample temperature decreased from 1473 K to room temperature, as shown in Figure S4, 5. Note that we could not distinguish individual boron and nitrogen sub-lattices from the single-shot HRTEM images recorded *in situ.* This is due to several reasons: 1) the too-close atomic number (Z) between boron and nitrogen ($\Delta Z=2$) and the charge redistribution in h-BN[34]; 2) the presence of residual aberrations, especially the three-fold aberrations, such as three-fold astigmatism $A_2$[35].

**GB Fabrication**. All the GBs mentioned in the main text were obtained *in-situ* at a sample temperature of 1073 K. Nanometer-sized holes were generated on the top and bottom layer in a bilayer h-BN via beam irradiation, whose edges are nitrogen terminated. When the two edges met, a junction spontaneously formed connecting the top and bottom layers locally, which further develops into a stable GB via subsequent structural relaxation. Notably, the electron beam irradiation area in HRTEM is larger than that recorded on the camera. Thus, GBs could also be formed outside the observing window whose formation processes were not captured. As a result, though we only conducted ~250 nano-welding attempts, more GBs (~430) were obtained.

Experimentally, keeping a high sample temperature was necessary for the success of nanowelding and the subsequent formation of GB and other interfaces, and it serves for a few purposes. i) To improve the sample cleanness via removing the surface contaminations and to depress the chemical etching as well[36]. For instance, the as-formed GBs can be easily damaged due to chemical etching effects induced by electron beam when the sample temperature is below 673K, and hence the success rate of nano-weldings will be reduced; ii) To exclude the possible carbon doping into lattice and the GBs of h-BN; iii) to retain the edge neat, i.e., free of surface contamination or passivation, and thus maintain the edge reactivity high, all of them favor the merging of two open edges. Previous work confirmed the hydrogen heavily passivated edges of growing h-BN domains during the CVD, and thus cause the domains to grow vertically form to an overlapping GBs, rather than planar GBs[10]; iv) to thermally activate the necessary reconstruction and relaxation of the as-formed GBs for transforming into thermal dynamically preferable configurations.

It should be noted that either the electron beam or thermal heating does not influence the control of θ, while introducing structural relaxation of the GB which leads to the change of φ. The variation of φ during relaxation is mainly determined by the energy difference between the facets with different φ, and this difference becomes quite remarkable when θ is large, e.g., 60°-GBs in Figure 1e, and thus results in apparent inclination change. However, for GB with smaller θ, e.g., 30°-GBs in Figure 1d, the change of φ is small. Note that an accuracy of 6° for φ was derived by considering the small variations of inclinations during the experiments, as discussed above (see parameter-space in Method Section).

Related to the stability of these as-formed GBs, they should be thermodynamically stable as evidenced by different aspects: First, e-beam irradiation alone should have limited effects on the resulting GB configurations. For example, despite the fact that the boron atoms are prone to be sputtered by the incident electron beam compared to nitrogen atoms, the GBs obtained experimentally still possess a higher density of B-B bonds rather than N-N bonds, as shown in Figure S18, S19. Moreover, the GB structures remained unchanged after 5-mins annealing in the absence of e-beam irradiation, as shown in Figure S21. Second, sample heating also has minimal influences on GB structures. We also conducted experiments at 673K and 298K, in which the conclusions on the GB structures remain the same, e.g., GB faceting still existed, as shown in Figure S21. Third, though e-beam and high temperature will active GB motion and slightly influence the detailed structures of some GBs, they would not affect the main conclusions on the registry of GBs with different |θ| and φ which were based on abundant data and statistics. Fourth, to eliminate the effects of e-beam and heating furthest, the atomic resolution HRTEM images of GBs displayed in this manuscript were captured when GBs got fully relaxed, i.e., after long-term annealing and keeping unchanged for certain period.

**Image processing**. HRTEM images were Wiener filtered[37] to increase the signal-to-noise ratio. The illumination variations in HRTEM images were also removed by the band-pass filter in Image J[38]. And the image contrasts were inverted to make the atoms bright for better interpreting. More details are shown in Figure S6. HRTEM simulations were conducted in MacTempas, with the input parameters the same with the experimental settings: spherical aberration C3=5 μm, defocus Δf=-5.6 nm and spread of defocus of 3 nm.

ASSOCIATED CONTENT

**Supporting Information**.
The following files are available free of charge.
More details on the experimental results (PDF)
Movies showing the formation of a 30°-GB and a 60°-GB (mp4)

AUTHOR INFORMATION

**Corresponding Author**

*E-mail. chhjin@zju.edu.cn

**Author Contributions**

The manuscript was written through the contributions of all authors. All authors have given approval to the final version of the manuscript.

**Funding Sources**

This work was financially supported by the National Natural Science Foundation of China under Grant No. 51772265, No. 51761165024 and No. 61721005, the National Basic Research Program of China under Grant No. 2015CB921004, the Zhejiang Provincial Natural Science Foundation under Grant No. D19E020002, and the 111 project under Grant No. B16042. The work on electron microscopy was done at the Center of Electron Microscopy of Zhejiang University.

**Notes**

The authors have no competing interests.

ACKNOWLEDGMENT

We thank Dr. Jidong Li and Prof. Wanlin Guo for providing high-quality h-BN films, and Profs. Jun Yuan and Ze Zhang for fruitful discussions and critical comments.

ABBREVIATIONS

GBs, grain boundaries; TEM, transmission electron microscope; h-BN, hexagonal boron nitride; APBs, anti-phase boundaries;

# Supporting Information

## Atomic-Precision Fabrication of Quasi-Full-Space Grain Boundaries in Two-Dimensional Hexagonal Boron Nitride

Xibiao Ren, Xiaowei Wang, Chuanhong Jin*

State Key Laboratory of Silicon Materials, School of Materials Science and Engineering Zhejiang University, Hangzhou, Zhejiang 310027, China

*Corresponding authors: chhjin@zju.edu.cn

**Table of contents**

# 1. Out-of-plane warping in the junction between the top and bottom layers

The height difference (~3.35 Å)[1] between the top and bottom layer introduces warp into the welded GB, especially at its two endpoints. We first discuss the impact of the strain field of two endpoints, taking a junction with the same orientation (θ=0°) as an example (no dislocations in the junction region). We then turn to discuss the tilted GBs.

## 1.1 Out-of-plane warping in the junction with the same orientation (θ=0°)

Figure S1 shows the quantitative analysis of a junction formed between the top and bottom layers, both of which share the same orientation ($\theta_t$=0°) while possessing a slight shift. As shown in Figure S1c, geometric phase analysis (GPA)[2] confirmed there exist two dislocation cores in the two endpoints, and deformations in the interlayer welded region (interlayer GB). As TEM images are the 2D projection of 3D structures, the dislocation cores might also be introduced by the additional lattice plane in the projected junction region as indicated in Figure S1d, g. Similar dislocation cores were also frequently observed in bi-layer 2D h-BN flakes with a non-zero interlayer twisting angle (Figure S2).

The height information usually gets lost in 2D-projection TEM images. In the regions away from the dislocation core, the bond length barely changes[3], and thus these regions can be treated as 'perfect' lattice. So deformations in these 'perfect' regions (in 2D projection) are caused by the inclination/warping of the 2D membrane. By simple geometrical analysis, we can derive quantitatively the distance-dependent height displacement and inclination angle from strain maps in GPA, as shown in Figure S1f-i. However, this method cannot distinguish the two different inclination directions (positive or negative). Fortunately, in the junction between the top and bottom layers, only one inclination direction (either positive or negative) is predominant, and thus the calculated height displacement should originate from this inclination. Under this situation, the existence of small inclination in the other direction and small compression due to the change of bond length will overestimate the magnitude of height change.

When the distance between two endpoints ($L_{ep}$) increases from ~3 nm (Figure S1b) to ~10 nm (Figure S1k), deformations between them decrease if one compares the results shown in Figure S1E-I with that in Figure S1o-r. As shown in Figure S1q, the mean inclination angle in line profile b (or c), corresponding to a 3.6 nm (or 2 nm) distance away from dislocation core is ~9.5° (10.9°). This value is comparable to that caused by the ripples in graphene, typically 5°~8° at room temperature[4,5]. And the height displacement is almost the same as the vicinity of dislocation in graphene (5−20 Å)[3]. Considering the overestimation in height displacement in the calculation and assuming a height difference of 3.35 Å (the interlayer distance in bilayer h-BN), we can deduce the mean inclination angle in line profile b or c (3.6 nm/2 nm away from dislocation core) to be ~6°, which has a similar magnitude to that in graphene ripples (5°~8° at room temperature).

To sum up this part, when $L_{ep}$ is large enough (for example, > 4nm), the out-of-plane warping caused by either the junction and/or the height difference is equivalent to the deformation resulted from two dislocation cores at two endpoints of the junction and the inherent ripple (with a mean inclination angle of 6-10°). We know that ripples extensively exist in 2D materials. Thus, the primary difference between (the interlayer) GBs obtained by the quasi-bicrystal approach and regular planar GBs lie in the deformation caused by the two dislocation cores at two endpoints.

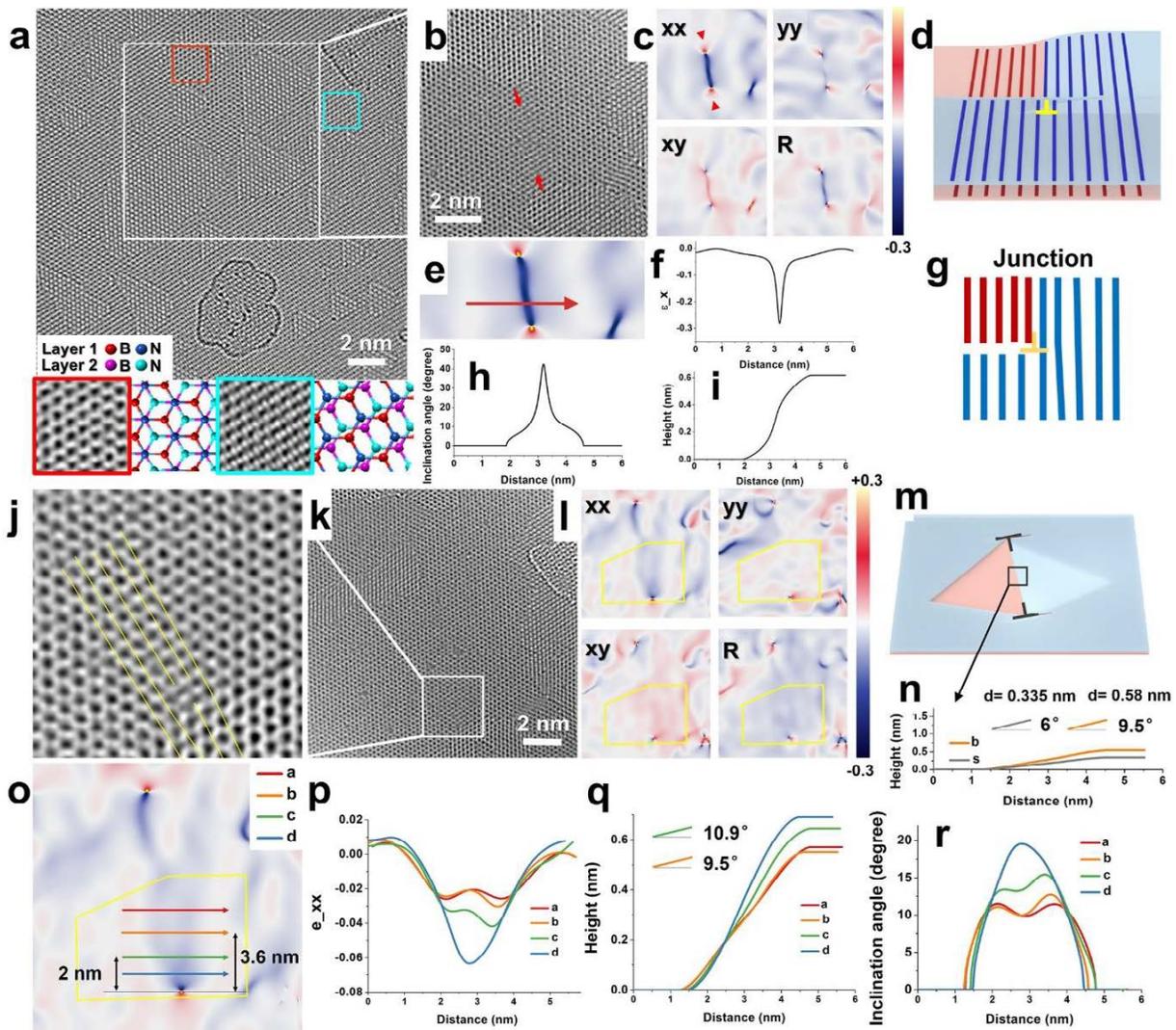

**Figure S1. The warping introduced due to the height difference between the top and bottom layers in bilayer h-BN. a,** Low magnification HRTEM image of this bilayer h-BN flake, where the top and

bottom layers have the same orientation but a slight shift from the AA stacking. And close-up inspection of the local stacking from the red and blue boxes are displayed in the bottom panel, respectively. **b,** An enlarged HRTEM image of the junction between two layers, and the corresponding GPA maps is displayed in **c**. The red arrows point out the two endpoints of the junction. "xx", "yy", "xy", and "R" in GPA maps represent $\varepsilon_{xx}$, $\varepsilon_{yy}$, $\varepsilon_{xy}$, and the rotation parts, respectively. **d,** A schematic diagram showing the introduction of a dislocation by the junction between the top and bottom layers, and the corresponding top-view is displayed in **g**. **e,** An enlarged GPA map of the $\varepsilon_{xx}$ in **c**. **f,** Line profile of $\varepsilon_{xx}$ in **e**. **h, i,** The calculated inclination angle and z-height based on by interpreting the variant of $\varepsilon_{xx}$. **k,** A HRTEM image of the junction region recorded a few minutes after its formation, and **j** showing the enlarged atomic-resolution HRTEM image. Yellow lines represent the (110) lattice planes. **l,** GPA maps for the HRTEM image shown in k. **o,** An enlarged GPA map of $\varepsilon_{xx}$ part. The yellow polygon indicates the region where $\varepsilon_{xx}$ is prominent. **p,** Line profiles of $\varepsilon_{xx}$ in o. **q,** Calculated z-heights and inclination angles by interpreting the variant of $\varepsilon_{xx}$. **m,** A schematic diagram illustrating the structure shown in **k**. **n,** Calculated z-heights (line b in **q**) versus the distance. x, y axis have equal scales, which can visually reveal the inflection. The mean inclination angles were calculated to be 6° and 9° respectively, assuming an interlayer distance d=0.335 nm and by the measured results.

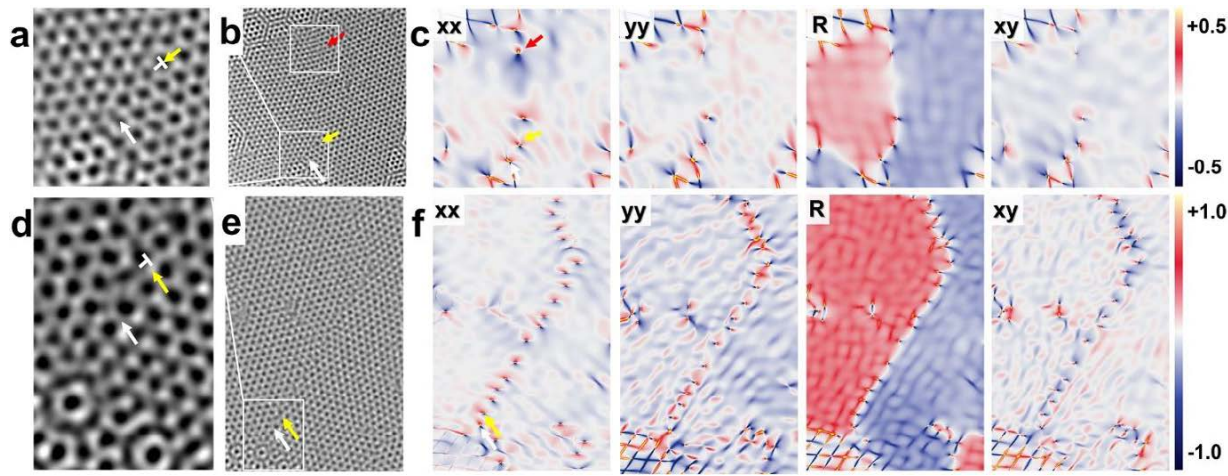

**Figure S2. The out of plane warping introduced by the height difference in low-angle and high angle GBs, respectively in 2D h-BN. a, b,** a 11°-GB, and **d, e,** a 27°-GB. Their corresponding GPA maps are displayed in **c** and **f**, respectively. "xx", "yy", "xy", and "R" in GPA maps represent $\varepsilon_{xx}$, $\varepsilon_{yy}$, $\varepsilon_{xy}$, and rotation parts, respectively.

**1.2 Out-of-plane warping in the tilted GBs obtained by our experiments**

We then discuss the warping of the tilted GBs. As discussed above, the major difference lying between the planar GBs and our nano-welded GBs can be attributed to the dislocation cores at the two endpoints of GBs. For small angle GBs as shown in Figure S2b, the distances between dislocation cores are large. It is seen that the dislocation/endpoint in the top region (marked by the red arrow), which was formed due to the junction of top and bottom layers, will influence the strain field in the GB region, as indicated in $\varepsilon_{xx}$ map in Figure S2c. But in the bottom region (white arrow in Figure S2a-c), the strain field of the dislocation at the endpoint was effectively canceled by the strain field due to the dislocation inside the GB (marked by the yellow arrow in Figure S2a-c), as indicated in $\varepsilon_{xx}$ map in Figure S2c. Such a strain field cancellation becomes even more pronounced for large-angle GBs in which the dislocations are closely packed, as shown in Figure S2d-f. As a result, the warping introduced by the dislocations at the endpoints of GB is easily canceled by the nearby strain field of the dislocations from the GB, and thus the impact of warping on the GB region is quite small.

To sum up, when the distance between two endpoints ($L_{ep}$) is large enough, in other words, the GB is long, and the strain field of end-point dislocations is canceled by the dislocations in the GB, our nano-welded GBs can be treated as planar GBs. All the GBs shown in Figure 3 in the main text possess a $L_{ep}$ > 5.5 nm, and mostly > 7 nm. Furthermore, the GBs displayed in the main text were in the selection region between two 'end dislocations'. The GB structures of low-angle sym-GBs confirm the results of theoretical prediction, which further proves that the GBs obtained can be treated as planar GBs.

**2. Out-of-plane warping in regular planar GBs**

The out-of-plane buckling should also be considered in planar GBs in 2D membranes[6,7], which will reduce the GB energy. The GB energy with respective to θ- $E(\theta)$ can be described by the divergent logarithmic term in Read–Shockley equation:

$$E(\theta) = \frac{\mu|\vec{b}|}{4\pi(1-\nu)} \theta \left(1 + ln\frac{|\vec{b}|}{2\pi r_0} - ln\theta\right) \quad (1)$$

since the long-range strain field of an isolated dislocation decays with distance as 1/r, where μ is the shear modulus, $\vec{b}$ is the Burger's vector, ν is the Poisson's ratio, and the core radius $r_0$ encompasses the dislocation energy. The logarithmic term of an isolated dislocation core can be safely omitted in the case of GBs in 2D materials, due to the offset of long-range strains caused by the out-of-plane deformation[8].

The out-of-plane warping in planar GBs becomes significant in small-angle planar GBs, which dramatically decreases with the misorientation angle $\theta^8$, as a result of effectively canceling of the strain fields of dislocations with opposite signs. Most of the GBs reported in this study have a misorientation angle |θ|>5°, which possess much smaller out-of-plane deformation than that for |θ|<5° GBs. Hence for simplicity, the out-of-plane deformations result from intrinsic/planar GBs are not considered here.

## 3. Fabrication of h-BN/graphene heterojunction

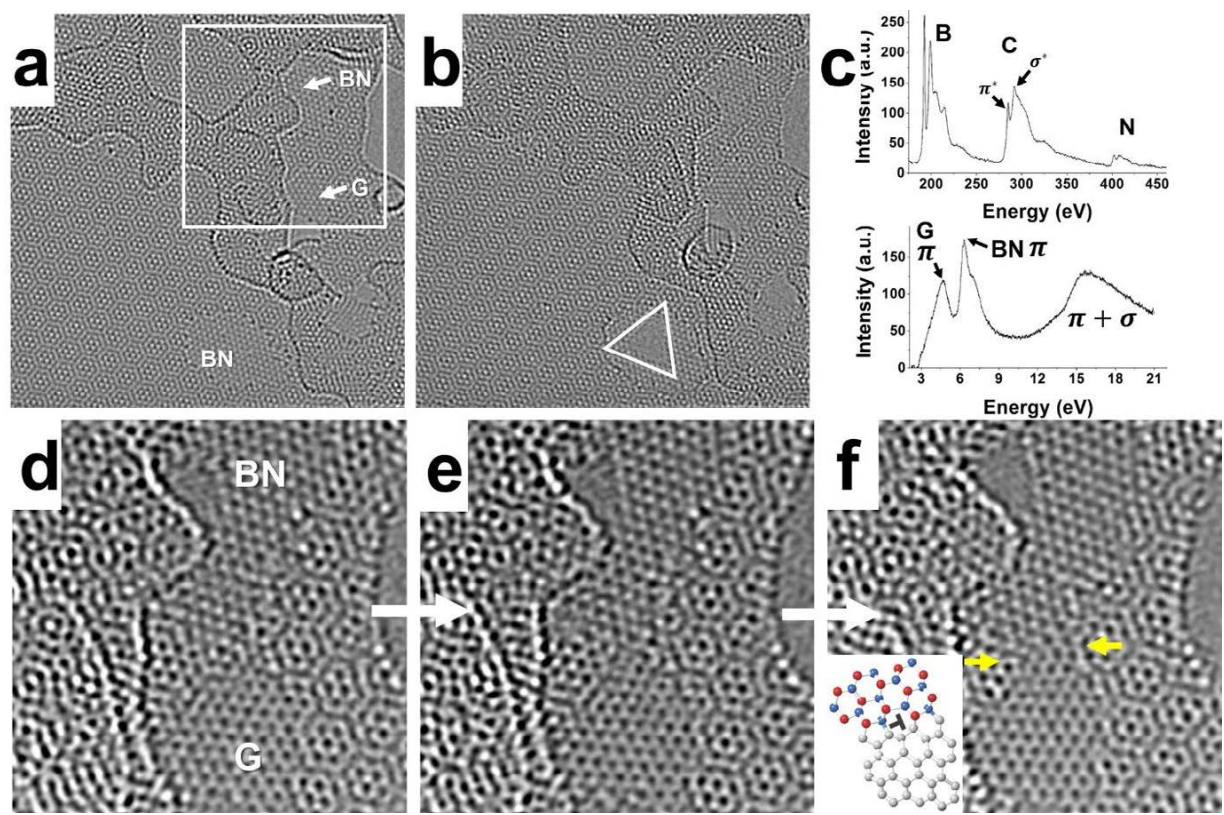

**Figure S3. Fabrication of a graphene-h-BN heterojunction. a, b,** Low-magnification HRTEM images of a h-BN/graphene vertical junction. A triangular hole was created in monolayer h-BN in **b**, while a circular hole was formed in the graphene flake as a result of metal atom assisted etching. **c,** Electron energy loss (EEL) spectrum of the h-BN and graphene (from a 150 nm×150 nm region). Top: the core-loss spectrum. Bottom: the low-loss spectrum showing the π and π+δ plasmon peaks of h-BN and graphene. **d-f,** Sequential HRTEM images showing the formation of interface (yellow arrowed) between graphene and h-BN.

## 4. Structural evolution of the in situ created holes in 2D h-BN under electron beam irradiation and sample heating

We first increased the sample temperature to 1473 K, under which the chemical etching of 2D h-BN will be largely depressed, and then gradually decreased the sample temperature and in the meantime monitored the associated structural evolution of holes. Importantly we found that the triangular holes well retain its crystallographic orientation under different sample temperatures during the whole process, through which was accompanied by slight shape variation, as shown in Figure S4, 5. As reported previously[9,10], these triangular holes are terminated with nitrogen atoms on the edges. Note that non-triangular holes reported by Thang et al.,[11] were rarely observed in this study, and they would eventually change to triangles as shown in Figure S4, S5.

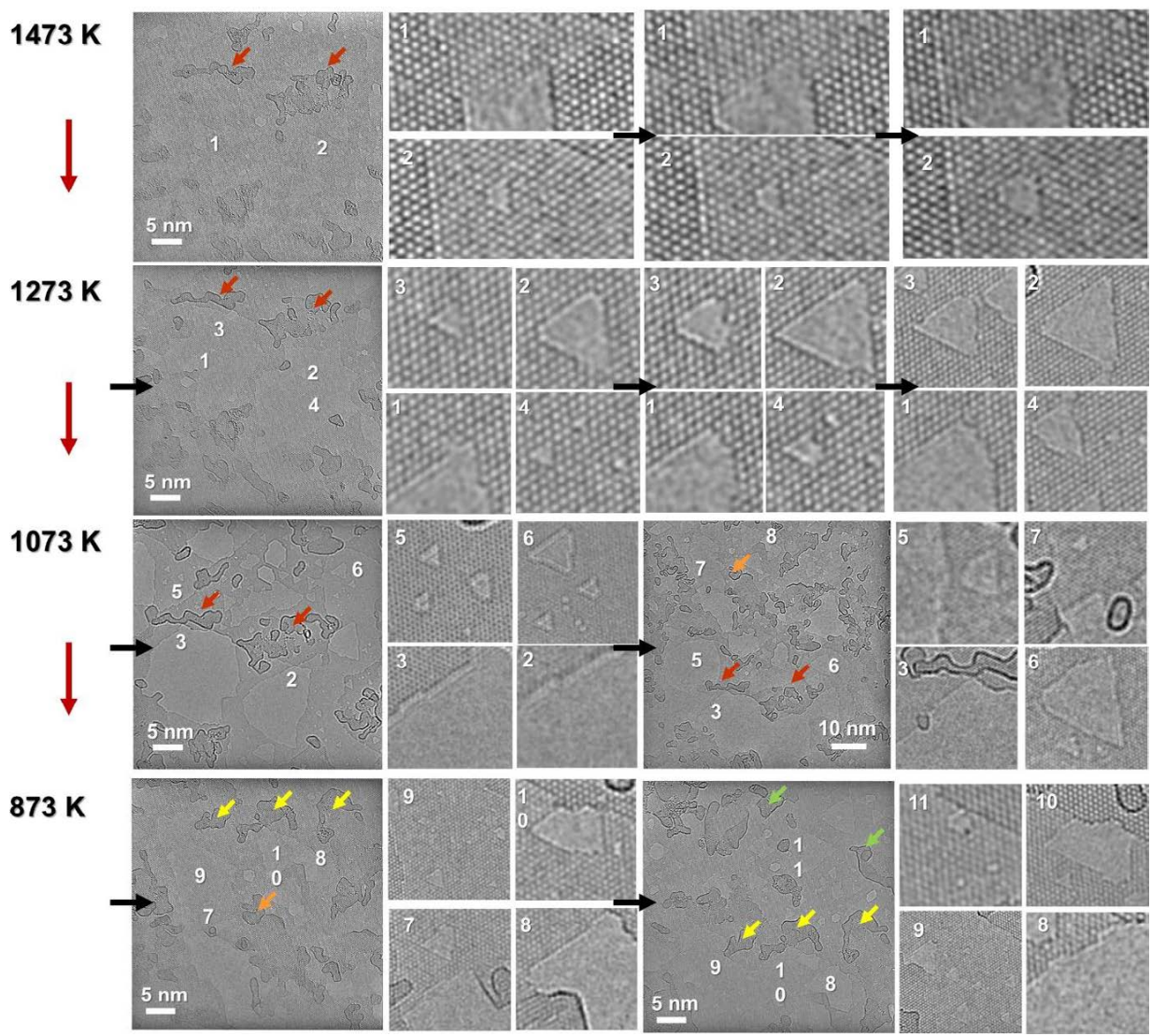

**Figure S4. The shape evolution of as-formed holes under different sample temperatures from 1473 K down to 873 K.** The black arrow marks the increasing of time. The skew arrows point out the presence of amorphous contaminations, which were used as markers. And the arrows in the same color in different

HRTEM images point to the same amorphous contaminations. The numbers-1, 2,…11 represent different hole-containing regions.

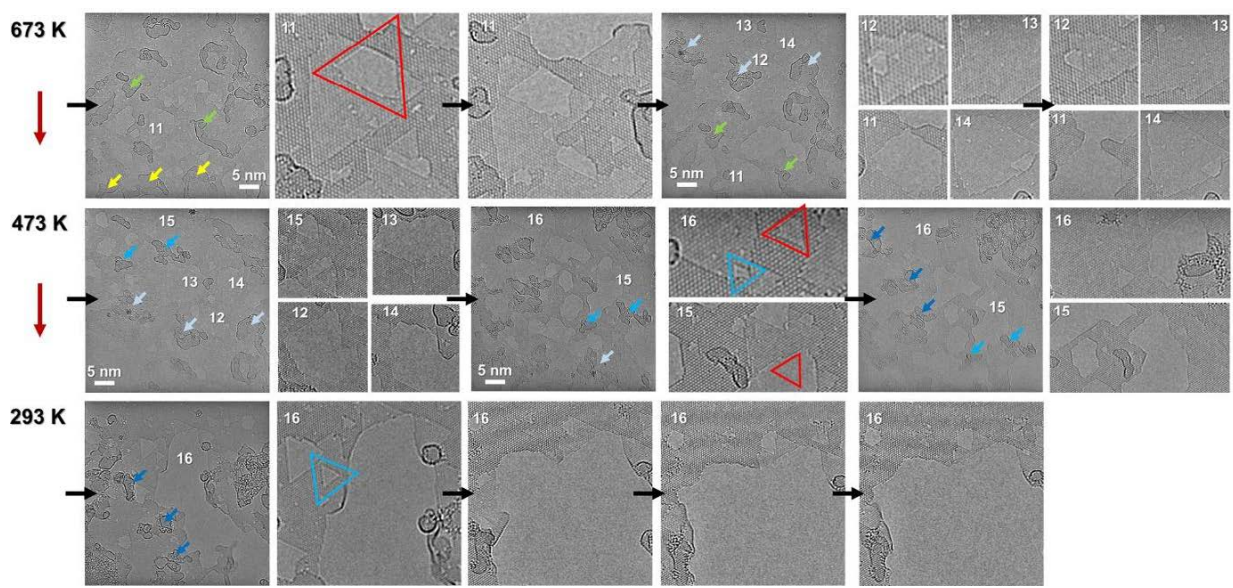

**Figure S5. The shape evolution of holes with the sample temperature changing from 673 K down to 293 K.** The sequential TEM image of 673 K was taken after the sample cools down from 873 K (Figure S4) while recorded in the same region (arrowed). Due to the loss of one layer with the orientation marked by red triangles in region 15 and 16, we use another layer (its orientation is marked by blue triangles in region 16) as a reference to see whether the hole shape change with temperature and time.

## 5. Imaging process

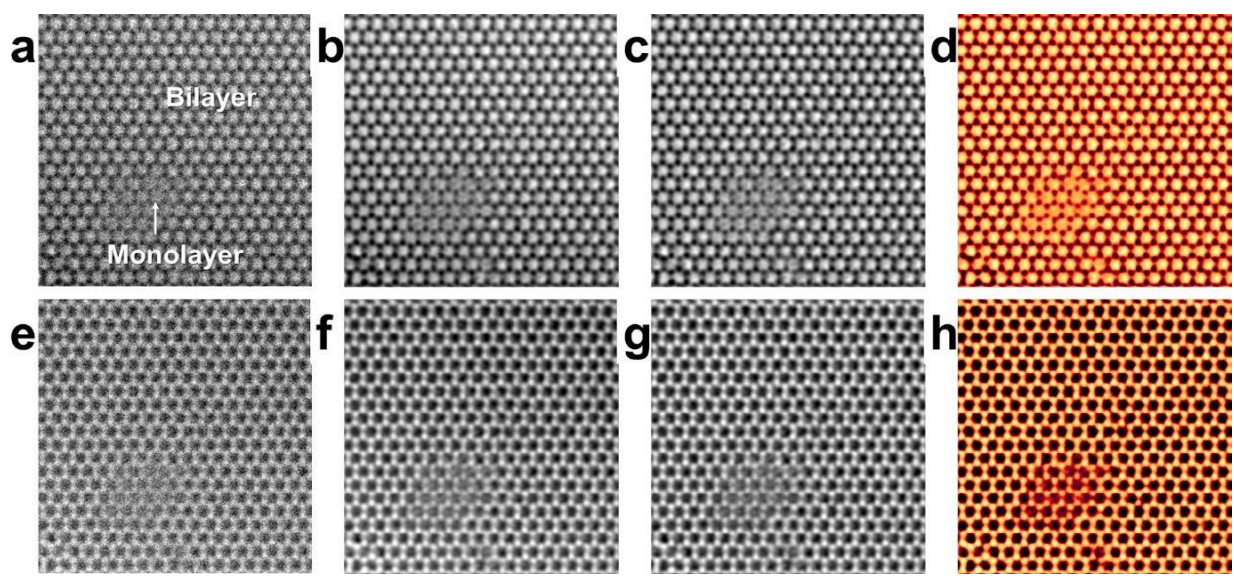

**Figure S6. Imaging process for the HRTEM from the perfect lattice from mono- and bi-layer h-BN. a,** A raw HRTEM image of bilayer and monolayer (in the hole). **b,** Filtered image. **c,** image after removing the illumination variation by the band-pass filter in Image J. **d,** image after applying fake color by Image J. **e-h,** inverted images of **a-d**, respectively.

## 6. Dislocations and line defects in 2D h-BN

Figure S7a-e illustrates five dislocation configurations and their corresponding Burgers vectors, respectively. A single disclination, resulting from an extra (missing) n×60° wedge in the lattice, has a strength of n×π/3, where n=1,2…., and introduces large elastic energy into the lattice. A positive and negative disclinations with the same strength pair form a dipole-dislocation[7,12], such as a pentagon-heptagon pair (5|7), or square-octagon pair (4|8). The dislocation strength is determined by inter-disclination separation, i.e., longer separation increases the strength, and vice versa, as indicated in Figure S7 a, b, d. In particular, the 585 shown in Figure S7f can be regarded as a pair of two opposite dislocations, resulting in a zero Burgers vector.

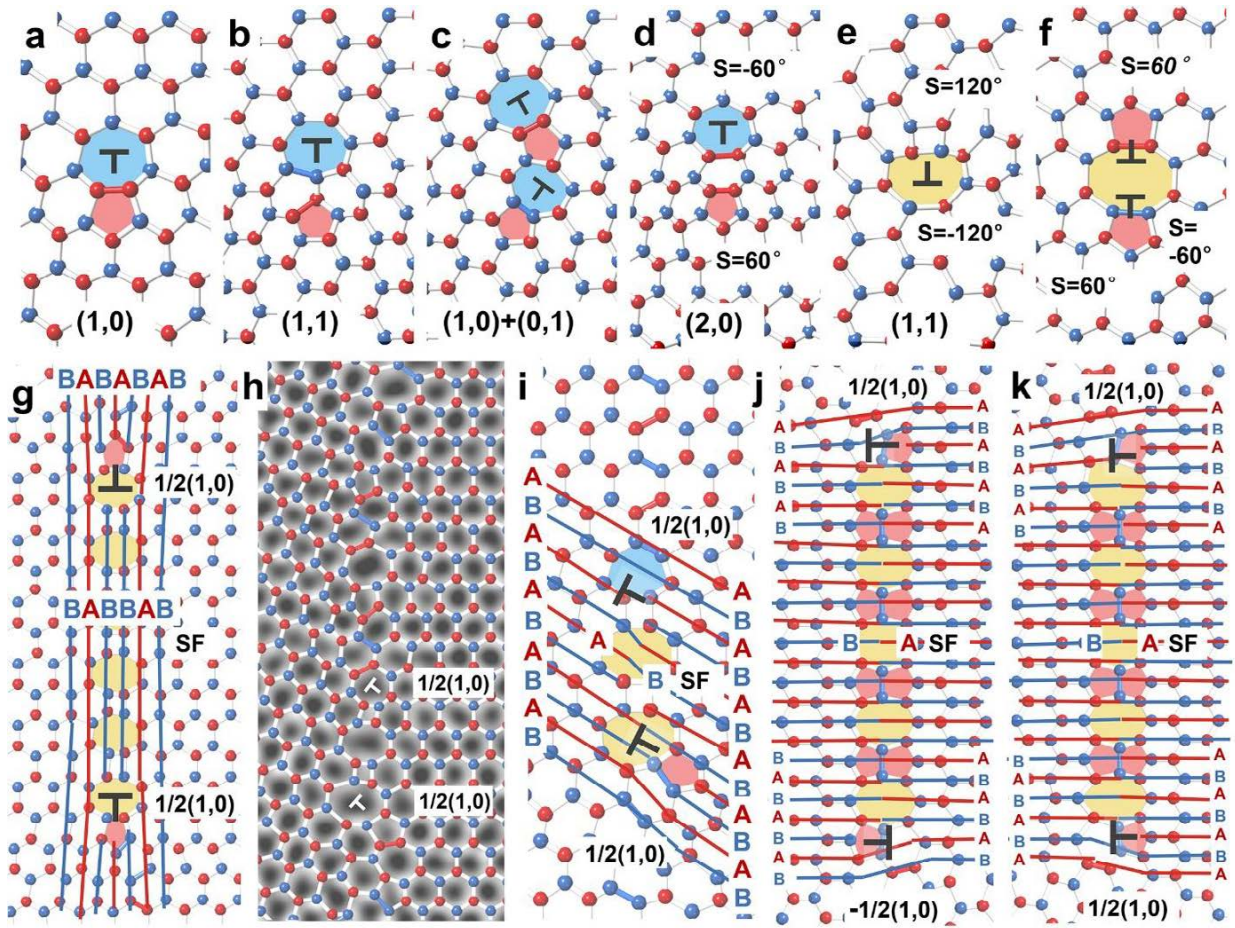

**Figure S7. Dislocations and line defects in 2D h-BN. a-e,** Five possible dislocations cores and **f** a 585 defect with zero Burger's vector in 2D h-BN. "S" represents disclination in **d-f**. B (boron) atoms: red, N (nitrogen) atoms: blue. **g,** A Frank partial dislocation loop. The red (A) and blue (B) lines represent (110) lattice planes. "SF" represents a stacking fault. **h, i,** Experimental results of the Shockley partial dislocation pair and its corresponding structure model, respectively. **j, k,** Shockley partial dislocations a 558s stacking fault.

We also experimentally observed another type of partial dislocation in high-angle asym-GB: Shockley partial dislocations as shown in Figure S7h. The corresponding structural model was displayed in Figure S7i. The Shockley partial dislocations contain two unidirectional Shockley partial dislocations-1/2(1,0)- and stacking faults-4|8s in between, which can also be understood via dislocation dissociation in which a (1,0) perfect dislocation breaks into two 1/2(1,0)s partial dislocations and a stacking fault in between. Figure S7g displayed a Frank partial dislocations loop with opposite signs. The stacking fault of partial dislocations can also be comprised of 558s as shown in Figure S7j, k, but was not found in our experiments.

## 7. Details for the full-space GBs

According to the coincidence site lattice (CSL) theory, the periodicities of two grains should be the same to prevent the accumulation of long-range elastic strains[13]. As shown in Figure S8a, the periodicities are defined as

$$l_i = a\sqrt{(m_i^2 + n_i^2 + m_i n_i)} \quad (i=1,2) \quad (2)$$

where a is the lattice constant (a=2.50 Å in 2D h-BN), $n_i$, $m_i$ (i=1,2) are integers, and ($n_1$, $m_1$) and ($n_2$, $m_2$) are matching vectors of the two domains, respectively. For instance, as shown in Figure S8a, a GB with $l_1=l_2=13a$ can also be labeled as (13,0)|(7,8), where $m_1=13$, $n_1=0$, $n_2=7$, $m_2=8$. The GB energy-$E(\theta,\varphi)$ usually has cusps at special $(\theta,\varphi)$ with high symmetry and low $\Sigma$, where $\Sigma=m_1^2+m_1 n_1+n_1^2$, the inverse density of coincidence sites[13].

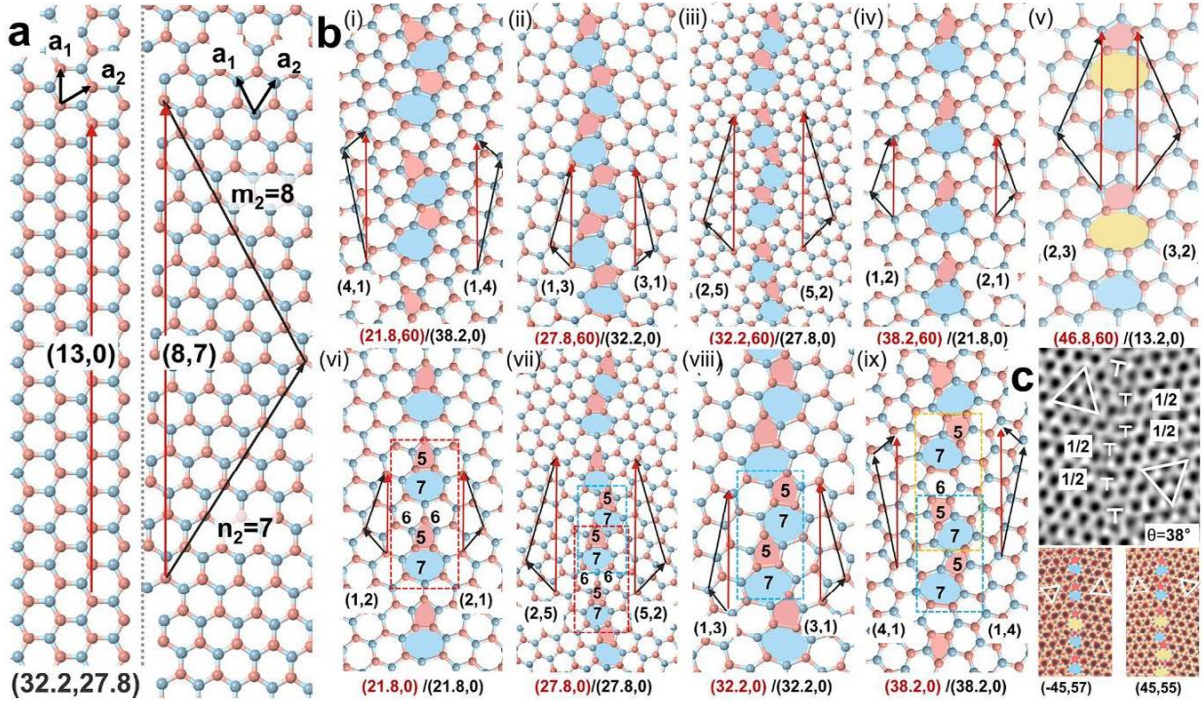

**Figure S8. The coincide site lattice (CSL) GBs in h-BN. a,** Definition of m, n and CSL GBs. The periodicity of each grain is described by two repeated vectors (13,0) and (8,7), respectively. **b,** Symmetric and anti-symmetric CSL GBs. Each GB was marked with two sets of symbols: one by the CSL theory (e.g., (4,1)), and the other in parameter space of binary lattice (red colors) (e.g., (21.8, 60)). **c,** 5|84|7s in anti-sym facet (θ=38°) and two anti-sym GBs (the same as that in Figure 3 in the main text).

Nine CSL GBs were displayed in Figure S8b, from which we observed that that anti-sym GBs with |θ|<38.2° comprise only 5|7s (Figure S8b (i-iv)); for anti-sym GBs with 38.2°<|θ|<46.8°, they comprise both 5|7s and Frank partial dislocations (Figure S8b (iv-v)), and for anti-sym GBs with 46.8°<|θ|<60°, they consist of only Frank partial dislocations, as indicated by Figure S8b (v). Experimental results also confirmed that Frank partial dislocations start to appear in anti-sym facet when |θ|≥38° as shown by Figure S8c. Due to the measurement errors, minor variations of the values of θ and φ are allowed. For example, 38.2° and 46.8° can be safely approximated to be 38° and 47°, respectively.

The atomic structures of GBs with |θ|<38°depend mainly on the strain energy, similar to that in graphene. As shown in Figure S11, the low-angle sym-GBs with misorientation angles |θ|<15° are composed of aligned (1,0) dislocations and with sole homoelemental bonding, either B-B or N-N. Inside the GB, the mean separation-D between dislocation cores accord well with Frank's equation as $D=b_{(1,0)}/(2\sin(|\theta|/2))^{14}$. On this basis, in sym-GBs the closest packing of (1,0) dislocation cores appears

when $|\theta|\sim21.8°^7$. Similarly, the (0,1) dislocation- with the other sole homoelemental bonding, either N-N or B-B are introduced, as found in a GB-(30°,0°) in Figure S11. If the difference in atomic species (B and N) was ignored, the definition of our GB parameter space predicted that GB-(60°-|θ|,60°-φ) and GB-(θ,φ) should be identical. Thus, (1,0)s and (0,1)s constructed the dislocation pairs ((1,0)+(0,1)) that forms the anti-sym GBs with $|\theta|<21.8°$, as exampled by the GB-(-11,58)/(49,2) and GB-(19,52)/(41,8) shown in Figure S11. However, there are also some differences in detailed structure between GBs ($|\theta|<38°$) of h-BN and that of graphene. For example, GBs with different φ possess dislocations with different homoelemetal bondings, as compare GB-(14,0)/(14,0) (only B-B bondings), GB-(20,31)/(40,29) (mainly B-B bondings) and GB-(19,52)/(41,8) (nearly equivalent density of B-B and N-N bondings).

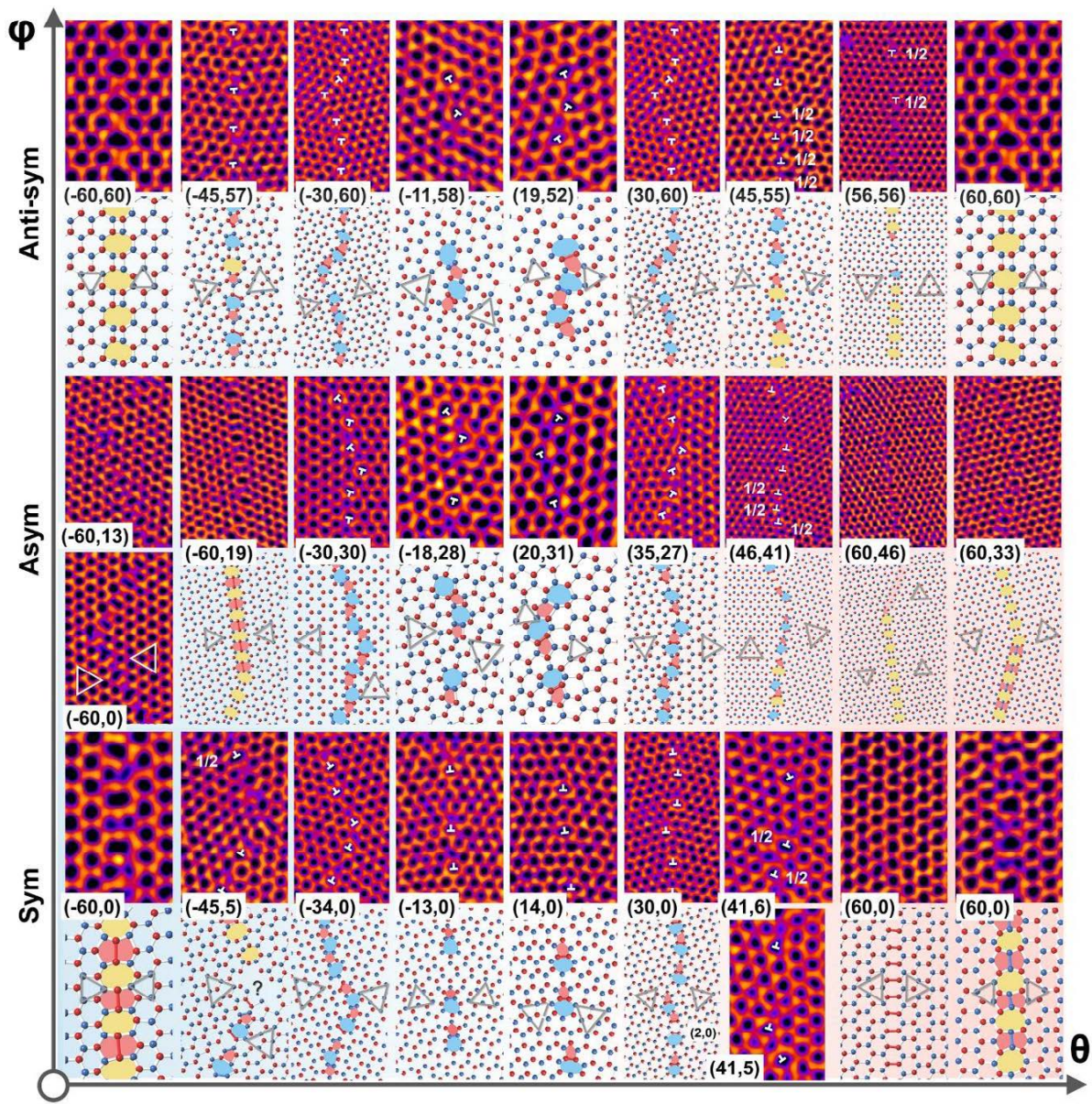

**Figure S9. The experimentally achieved quasi-full-parameter space GBs in 2D h-BN, which is the same as that in the main text (Figure 3).** Top: Experimental HRTEM images of the GB, bottom: the corresponding structural model. Pentagons, heptagons, and octagons in topological defects are filled with red, blue and orange, respectively. (-60,0) with two 48 facets and another |θ|>38° sym-GBs-GB-(41,5) is also displaced.

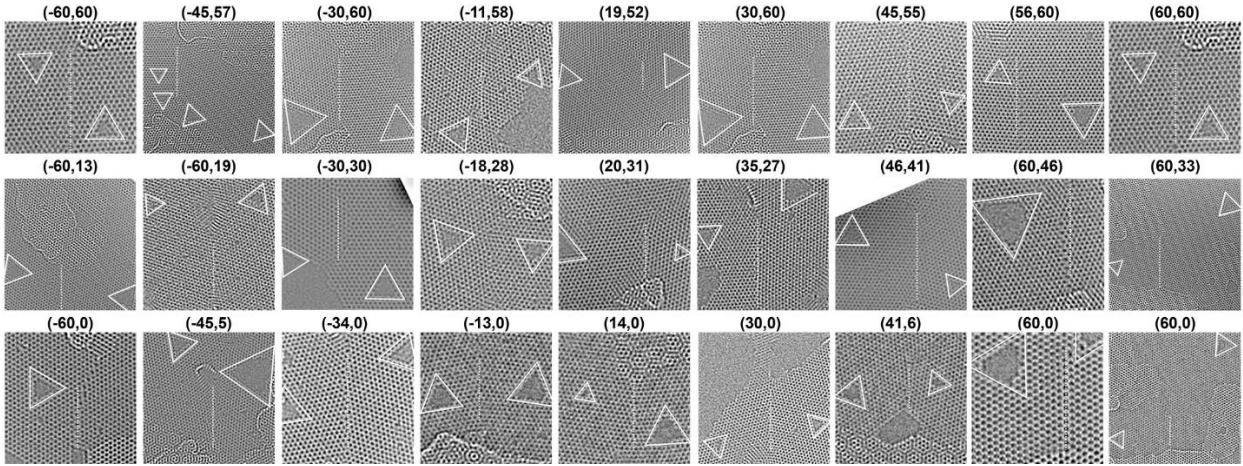

**Figure S10. Low-mag TEM images of the GBs shown in Figure 3 in the main text.** The grain orientation of the GBs is marked in the white triangle and the GB location is pointed out by the dashed white lines.

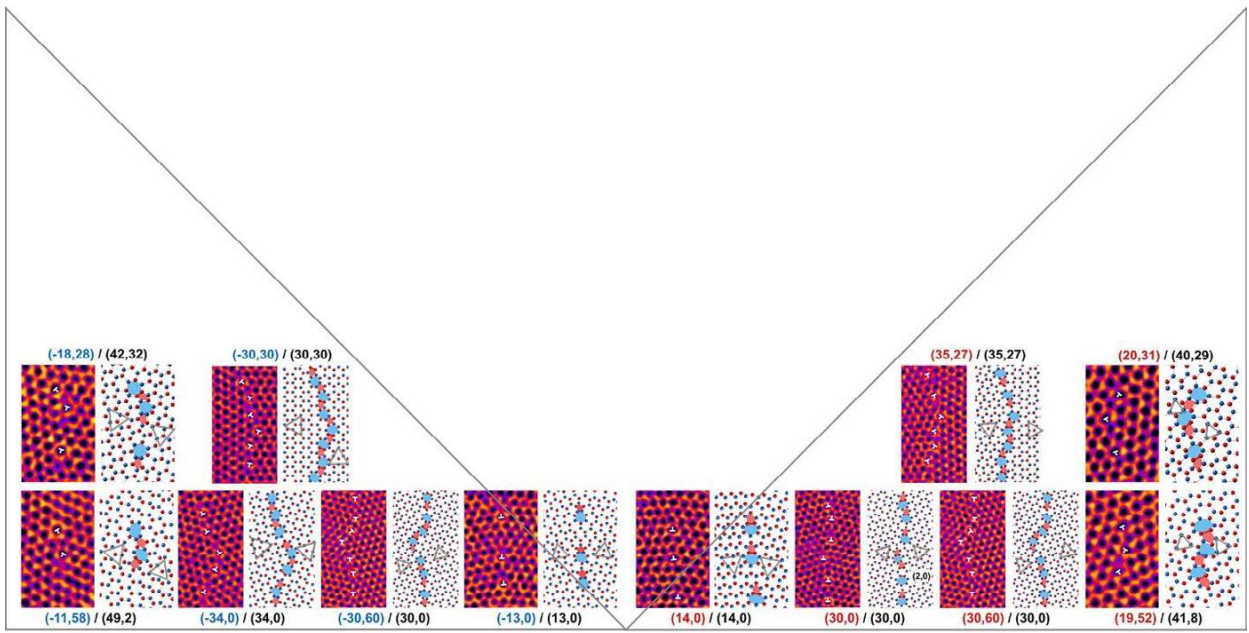

**Figure S11. |θ|<38° GBs in 2D h-BN.** The GB structures are the same as those shown in Figure 3 in the main text, but arranged differently-following the space description of GBs in graphene as marked by large grey triangles. Each GB was marked with two sets of symbols, e.g., (30,60)/(30,0): the former one (red and blue front) is described according to the space description of binary lattice like 2D h-BN, while the latter one is based on the space description of elemental hexagonal lattice like graphene.

## 8. The detailed structure of GBs with 66-BBs

As shown in Figure S12, a GB with 66-BBs does not contain any topological defects, which makes it difficult to be distinguishable from the pristine lattice of 2D h-BN. In practice, we proposed two methods to solve this problem relying on the fact that the bond length of B-B bonds in a 66-BBs GB and B-N bonds in pristine lattice of BN is different. First, the contrast of 66-BBs GB can be enhanced by proper image processing as we demonstrated in Figure S12g, H, during which most of the high-frequency lattice information was filtered out and only the low-frequency information was kept. Second, changing the defocus deviate from the Scherzer condition could also enhance the low-frequency information of a GB with 66-BBs; see the line profiles in Figure S12i. The contrast difference between a GB with 66-BBs and the pristine lattice in 2D h-BN are mainly caused by the difference in bond length between B-B (in a GB) and B-N (in pristine lattice) as indicated by Figure S12M, N, O. The consistency with the simulated HRTEM images further validates our statement.

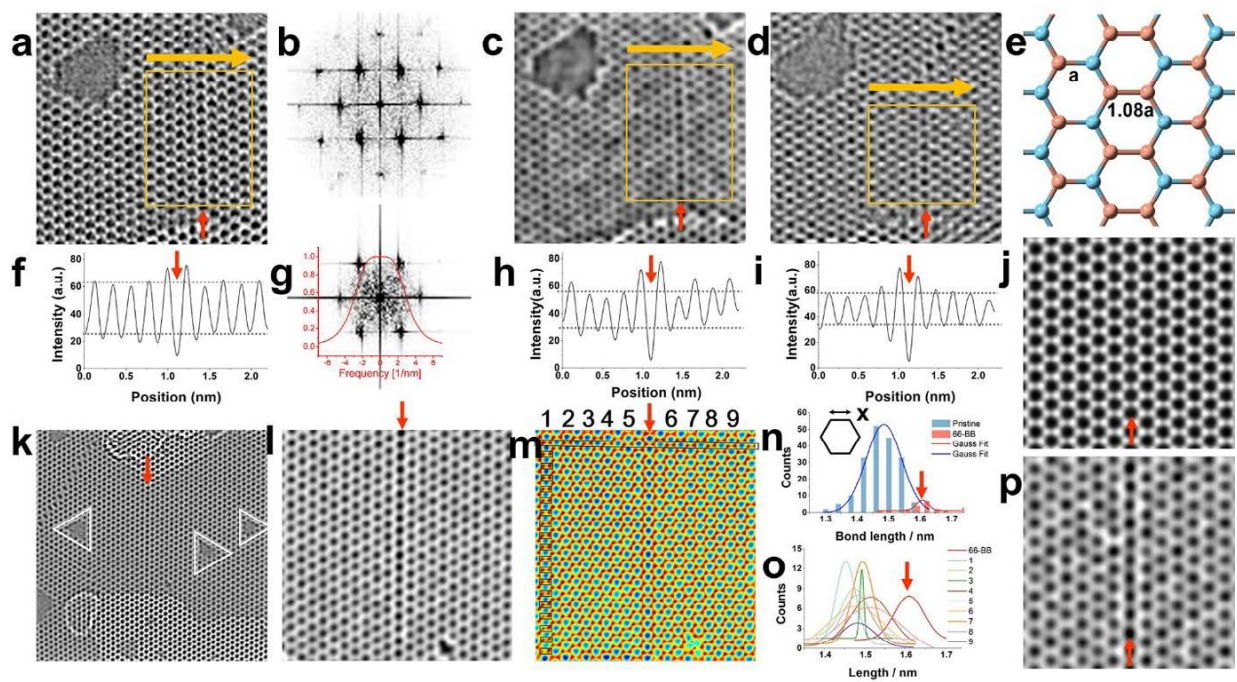

**Figure S12. Detailed atomic structure of 66-BB APBs. a, c, d,** An original HRTEM (corresponding FFT in B), filtered image (by the mask as shown in g) and HRTEM image (with large defocus value) of 66-BB, respectively. The red arrows indicate the position of the line defect. Their intensity line profiles are displayed in **f , h, i**, respectively, with their direction and width marked by orange arrows and rectangles in **a, c, d**. **e, j, p** Structure model of 66-BB ("a" represents the bond length of B-N bonding), corresponding simulated image (inverted) and its filtered image (by the mask as shown in g), respectively. **k,** A low-magnification image of 66-BB. **l,** Filtered image by the mask in G. **m,** An enlarged image of 66-

BB from K. **n,** Distribution of bond lengths of 66-BB and pristine h-BN in x-axis as indicate in M. **o,** Distribution of bond lengths of 66-BB and pristine h-BN of each row as numbered in k (x-axis).

## 9. The GB energy in 2D h-BN

In graphene, the GB energy of sym-GBs ($\varphi=0°$) is described by a generalized Read-Shockley equation[8],

$$E(\theta) = \frac{\mu|\vec{b}|}{4\pi(1-\nu)} |\sin(3\theta)| \times (\sum_{i=2}^{n} p_i \cos(3i\theta) + \sum_{i=1}^{nc} a_i |\cos(3\theta) - \cos(3\theta^{c_i})|) \quad (3)$$

where $\mu$=325.68 GPa is the shear modulus, $\nu$=0.318 is the Poisson's ratio, $p_i$, $a_i$ are dimensionless fitting parameters and the best-fiting gives $p_2$=-0.037, $p_3$=0.00618, $p_4$=0.0199, $a_1$=0.0891, $a_2$=0.2, respectively[8]. And the fitted results of the GB energy were shown by the black line in Figure S13a which exhibits an "M" dependence with the misorientation angle $\theta$.

Differently, the GB energy $E_f$ in 2D h-BN relates to the excessive elastic energy and the rising in chemical energy caused by the homoelemental bonding (B-B, or N-N), which can be estimated as $E_f = E_{el} + (n_{BB}\varepsilon_{BB} + n_{NN}\varepsilon_{NN} + n_B\mu_B + n_N\mu_N)/L$, in which $E_{el}$ is the strain energy per unit length of the GB, $n_{BB}$ ($n_{NN}$), $\varepsilon_{BB}$ ($\varepsilon_{NN}$) are the number, energy of B-B(N-N) bonds in a GB, respectively, $n_B(n_N)$ is the atom-number difference of B (N) atoms) between the examined and reference structures, $\mu_B$ ($\mu_N$) is the chemical potential, L is the length of a GB[15,16]. For B-rich sym-GBs (or the sym-facets), suppose they constitute 5|7s only as displayed in Figure S8b (vi)-(ix), the density of B-B bonding ($n_{BB}/L$) is derived as:

$$n_{BB}/L = \frac{1}{|b_{(1,0)}|/(2\sin\frac{\theta}{2})} \text{ for } 0°<\theta<21.8°, \quad (4)$$

$$n_{BB}/L = \frac{\sin(30°-\frac{\theta}{2})+\sin(30°+\frac{\theta}{2})}{a \times \sin(120°)} - \frac{2}{|b_{(1,1)}|/(2\sin(30°-\frac{\theta}{2}))} \text{ for } 32.2°\leq\theta\leq60°, \quad (5)$$

where a=2.50 Å is the lattice constant. Note that $n_{BB}/L$ equals to the dislocation density since each additional dislocation introduces a B-B bond for sym-GBs with 0°<$\theta$<21.8°. For $\theta$>32.2° sym-GBs, each additional dislocation (1,0)+(0,1) pair introduces two B-N bonds, as predicted by Frank's equation. Differently, for GBs with 21.8°<$\theta$<32.2°, the $\theta$ dependent of B-B density and GB energy becomes too complex, which were then numerically represented by the dotted lines for clarity, as shown in Figure S13a. Finally, the misorientation angle-$\theta$ dependent GB energy $E_f$ were quantitatively derived from the red lines in Figure S13a, b, which is also well consistent with the DFT results reported previously[17].

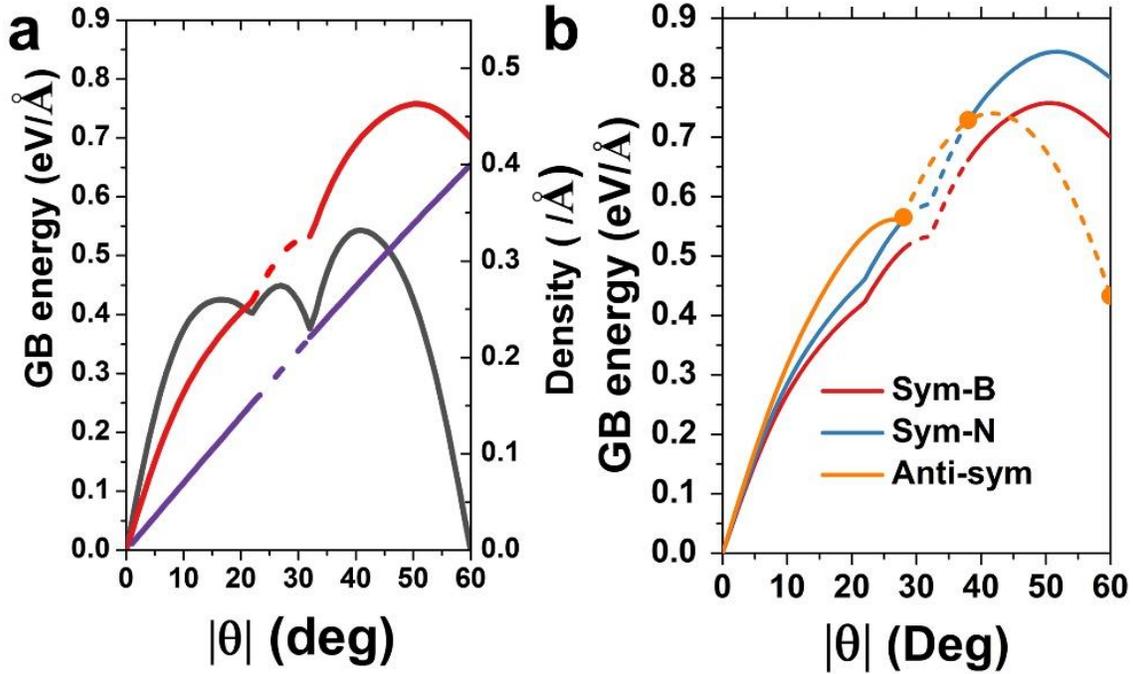

**Figure S13. GB energy in 2D h-BN. a,** Formation energies of GBs in graphene (black line), h-BN (red line), and the homoelemental bonding density of h-BN versus the misorientation angle θ (purple line, right vertical axis), respectively. **b,** Formation energies of B-rich symmetric, N-rich symmetric and anti-symmetric GBs in h-BN as a function of |θ|.

Similarly, for N-rich sym-GB, the θ dependent GB energy $E_f$ can also be derived as shown in Figure S13b. It should be addressed that influence of the chemical potential of the whole system was not considered, and the B-rich (N-rich) indicates the local chemical circumstance due to the presence of enriched B-B (N-N) bonds inside the GB.

Similarly, we could also derive the GB energy of anti-sym GBs as plotted in Figure S13b. Assuming the anti-sym GBs with |θ|≤27.8° constitute (1,0)+(0,1) dislocation pairs as displayed in Figure S8b (i)-(v), the density of homoelemental bondings $(n_{BB} + n_{NN})/L$ can be derived as:

$$(n_{BB} + n_{NN})/L = \frac{2}{(b_{(1,1)}/2sin(\frac{\theta}{2}))} \quad (6)$$

Note that an anti-sym GBs contains equal number of B-B and N-N bond, because each additional dislocation (1,0)+(0,1) introduces a B-B bond and a N-N bond as well. In particular, an anti-sym GB with |θ|=38.2° consists of only (1,0)s, and the density of homoelemental bondings $(n_{BB} + n_{NN})/L$ can be obtained as

$$(n_{BB} + n_{NN})/L = 1/(|b_{(1,0)}|/(2\sin(\frac{60°-38.2°}{2}))) \quad (7)$$

For anti-sym GBs with 27.8°<|θ|<38.2° and 38.2°<|θ|<60°, the |θ| dependent GB energy was numerically represented by the dotted lines for clarity, as shown in Figure S13b. Note that the formation energy of |θ|=60° anti-sym GB was adopted from previous DFT results[17].

## 10. More details about GB faceting

### 10.1 General description of GB faceting and the associated statistics

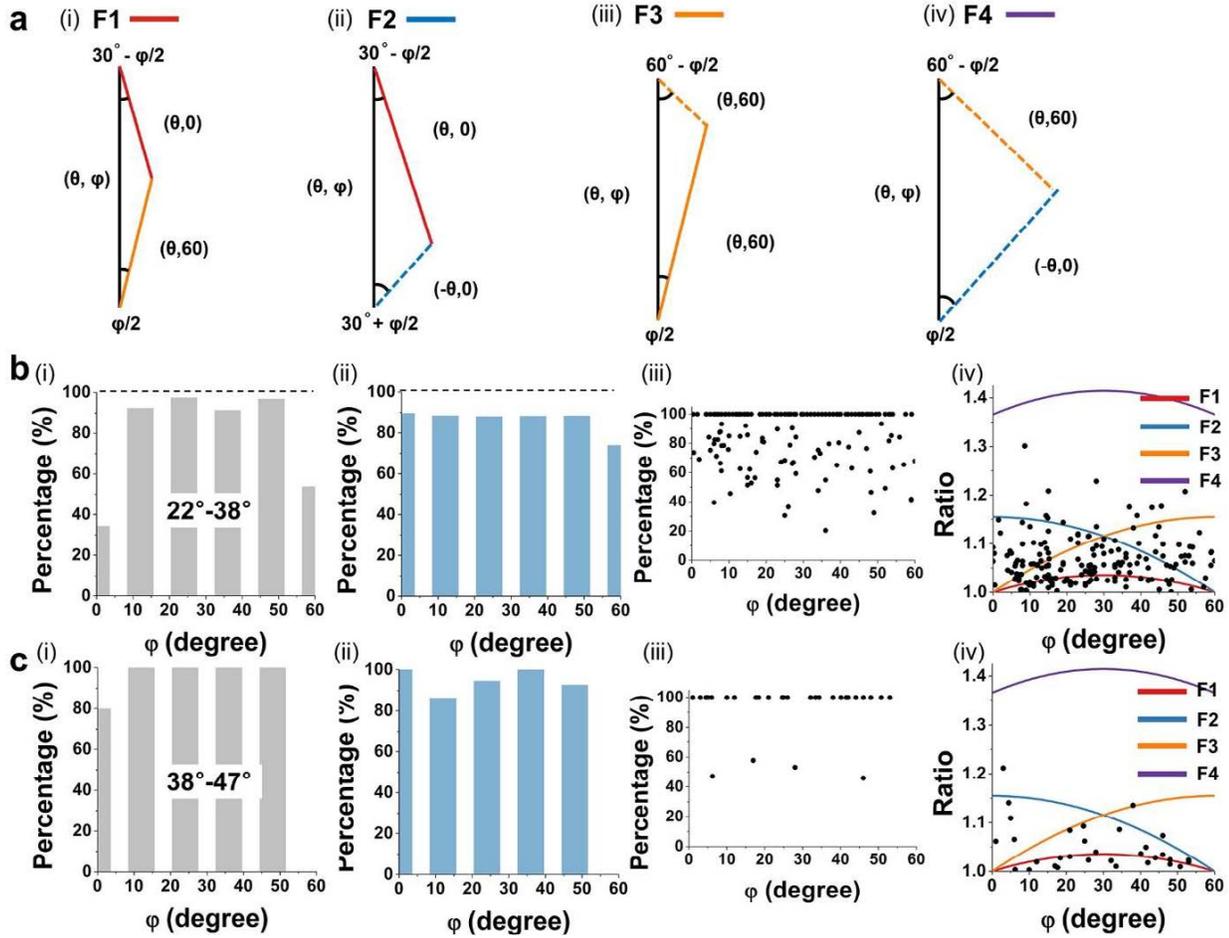

**Figure S14. More details on the statistics of GB faceting. a,** Schematic diagrams of four possible GB faceting. **b, c,** Statistics of faceting in |θ|=22°-38°and |θ|=38°-47° GBs, respectively. (i) The proportion of faceted GBs to all the GBs; (ii, iii) The average percentage and percentage of the length of symmetric/anti-symmetric facets to the overall length of the GB in faceted GBs; (iv) the ratio of the overall length of GB to the length of linear GB in faceted GBs, as a function of φ, respectively F1, F2, F3, F4 represent the ratios of GB faceting in A (i),(ii),(iii),(iv), respectively.

Detailed statistics on analyzing the type, length and their relative population were displayed in Figure S14. The proportion of the length of component symmetric/anti-symmetric facets inside a GB to the overall length of a GB is calculated as:

$$\frac{\sum_i ls_i}{\sum_i l_i} \times 100\% \quad (8)$$

where i represents a certain type of facet i, $ls_i$ is the length of a symmetric facet i within a GB, and $l_i$ is the length of each facet i within a GB. Here, we divide φ into seven ranges: 0°-6° for sym-facets, 6°-18°, 18°-30°, 30°-48°, 48°-54° for asym-facets, and 54°-60° for anti-sym facets, respectively. Then, in each range the average proportion of the length of symmetric/anti-symmetric facets to the overall length of a GB can be further written as:

$$\frac{\sum_{ij} ls_{ij}}{\sum_{ij} l_{ij}} \times 100\% \quad (9)$$

where $ls_{ij}$ is the length of a symmetric/anti-symmetric facet i along the GB j, and $l_{ij}$ is the length of each facet i along the GB j.

To minimize total GB length, a GBs only break into nearby facets, i.e., a GB-(θ,φ) breaks into (θ,0)+(θ,60), or (θ,0)+(-θ,0), or (θ,60)+(θ,60) in Figure 14a, as evidenced by the statistics in Figure 14b, c (iv) where $\sum_i l_i / l_0$ mostly falls into the regime that corresponding to the three faceting with shortest GB length in Figure 14a (i-iii). $\sum_i l_i / l_0$ is the ratio of the total length of experimental GBs to that of the linear GBs. For the faceting shown Figure S14a (iv), the length of overall component facets is too large, resulting in high formation energy. Thus this faceting type is unlikely happen in the experiments.

GB faceting in Figure S15a can be interpreted by the parameter space of GBs, as shown in Figure S15b, where a asymmetric GB-(θ,φ) breaks into two facets: a symmetric GB-(θ,0) and a symmetric GB-(-θ,0) by varying the φ in two opposite directions. Two facets have smaller formation energy than that of asymmetric GB-(θ,φ) as indicated by Figure S15c. GB faceting in sym-GBs can also be interpreted in a parameter space as shown in Figure S15d, e.

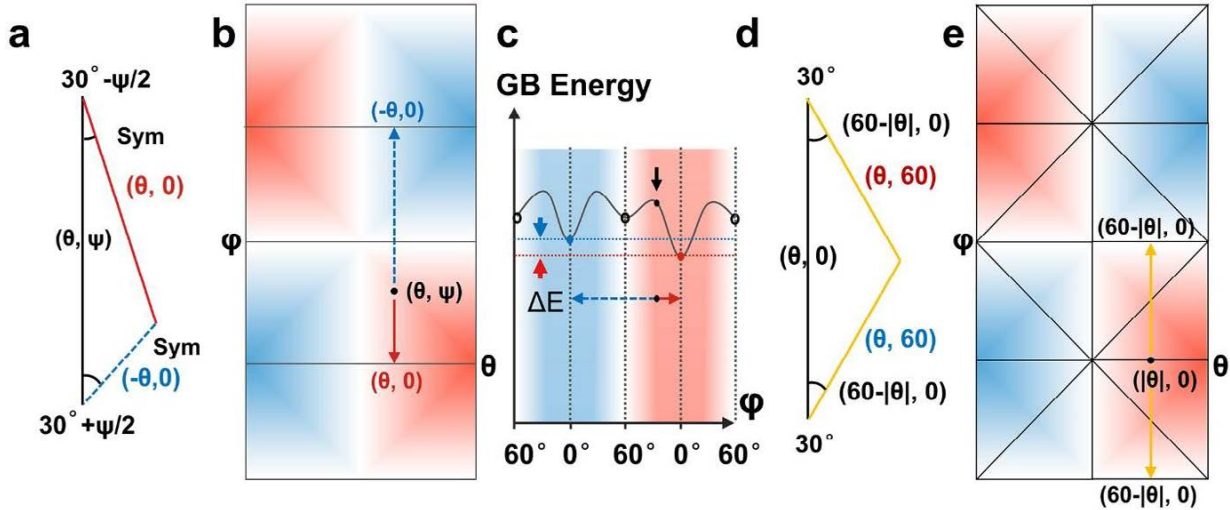

**Figure S15. GB faceting in 2D binary lattice. a, d,** Schematic diagrams showing the faceting in asym-GB and sym-GB, respectively. **b, e,** Periodic parameter spaces of GBs in 2D binary lattice, in which the triangle frames represent the space of graphene. **c,** Schematic diagram of GB energy as a function of φ.

### 10.2 More details about GB faceting in |θ|<38° GBs

In sym-GBs with an arbitrary θ- non-CSL GBs, the dislocation cores are not well separated along the GB line. As a result, the strain field of dislocations will concentrate locally; see red arrowed regions in Figure S16b. To minimize the strain energy, it is desirable to introduce another strain field reversing to the existing one by adjusting the location of dislocations (Figure S16a), or through the GB faceting, see the blue arrowed regions in Figure S16b and Figure S16c.

More examples showing the GB faceting (|θ|<38°) are displayed in Figure S17. For an asym-GB with a small φ, or equivalently 60°-φ, it consists of one dominant facet and other shorter facets (or kinks), as indicated in Figure S17c. Moreover, 5|84|7 facets likely to appear in the anti-sym with an inclination φ~38°, as shown in Figure S17d.

As summarized in Figure S18 a, b, there exist three types of faceting which possess different symmetric B-rich, symmetric N-rich length proportion. Thus, by comparing the symmetric B-rich length proportion in θ>0° GBs and symmetric N-rich length proportion in θ<0° GBs (with the same φ as that of θ>0° GBs), we can again determine the relative stability of B-B and N-N bondings. The definitions of length proportion and averaged length proportion in Figure S18 were the same as that in Figure S14. If the formation energy is identical between B-B and N-N bonding, the red lines (θ>0° GBs) should overlap the blue circles (θ>0° GBs) in Figure S18 d, f. However in reality, we observed the separation between these two data sets, especially when φ is small (the polarity of GB is large, either boron dominant or nitrogen dominant). And θ<0° GBs with small φ are mainly composed by anti-symmetric facets, as evidenced by Figure S18f. The asymmetric faceting behavior indicates that B-B bonding is more stable than N-N bonding, which is also consistent with theoretical calculations[17].

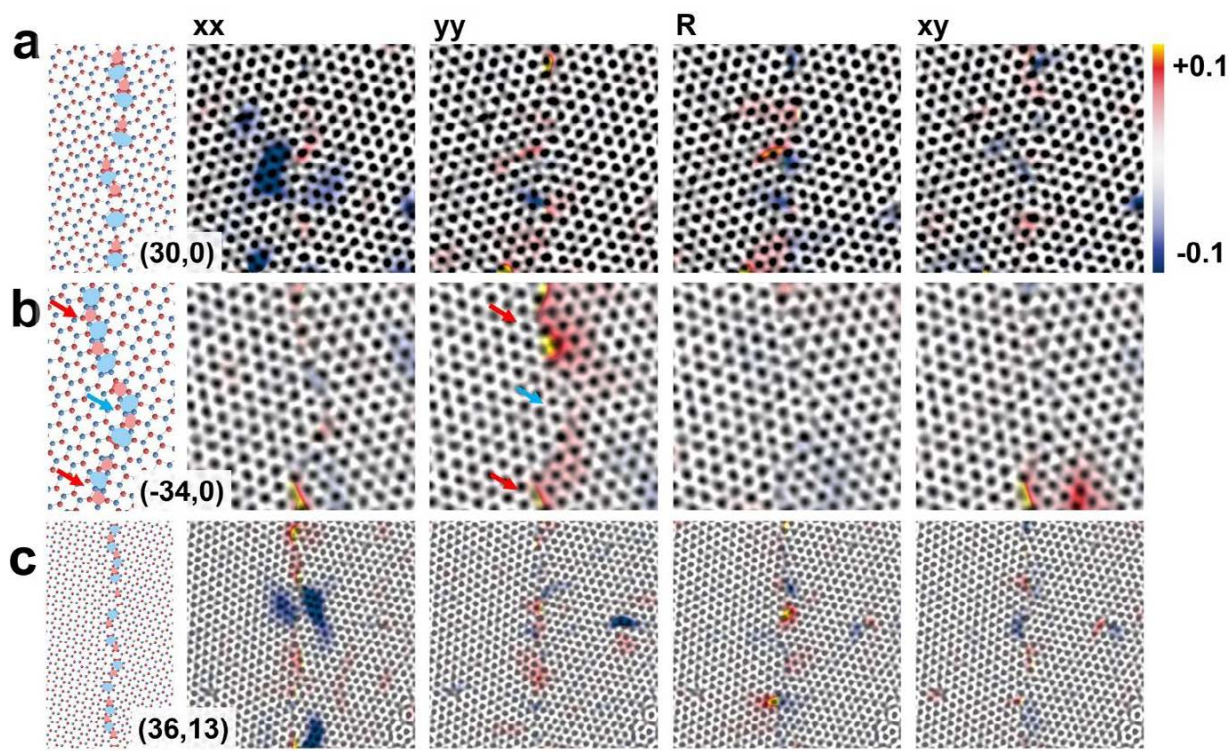

**Figure S16. Strain field of GBs. a-c,** Structure models and their strain field of GB-(30,0) in Figure 3 and GB-(-34,0) in Figure 4 in the main text, and GB-(36,13) in Figure S17. "xx", "yy", "xy", and "R" in GPA maps represent $\varepsilon_{xx}$, $\varepsilon_{yy}$, $\varepsilon_{xy}$, and rotation parts, respectively.

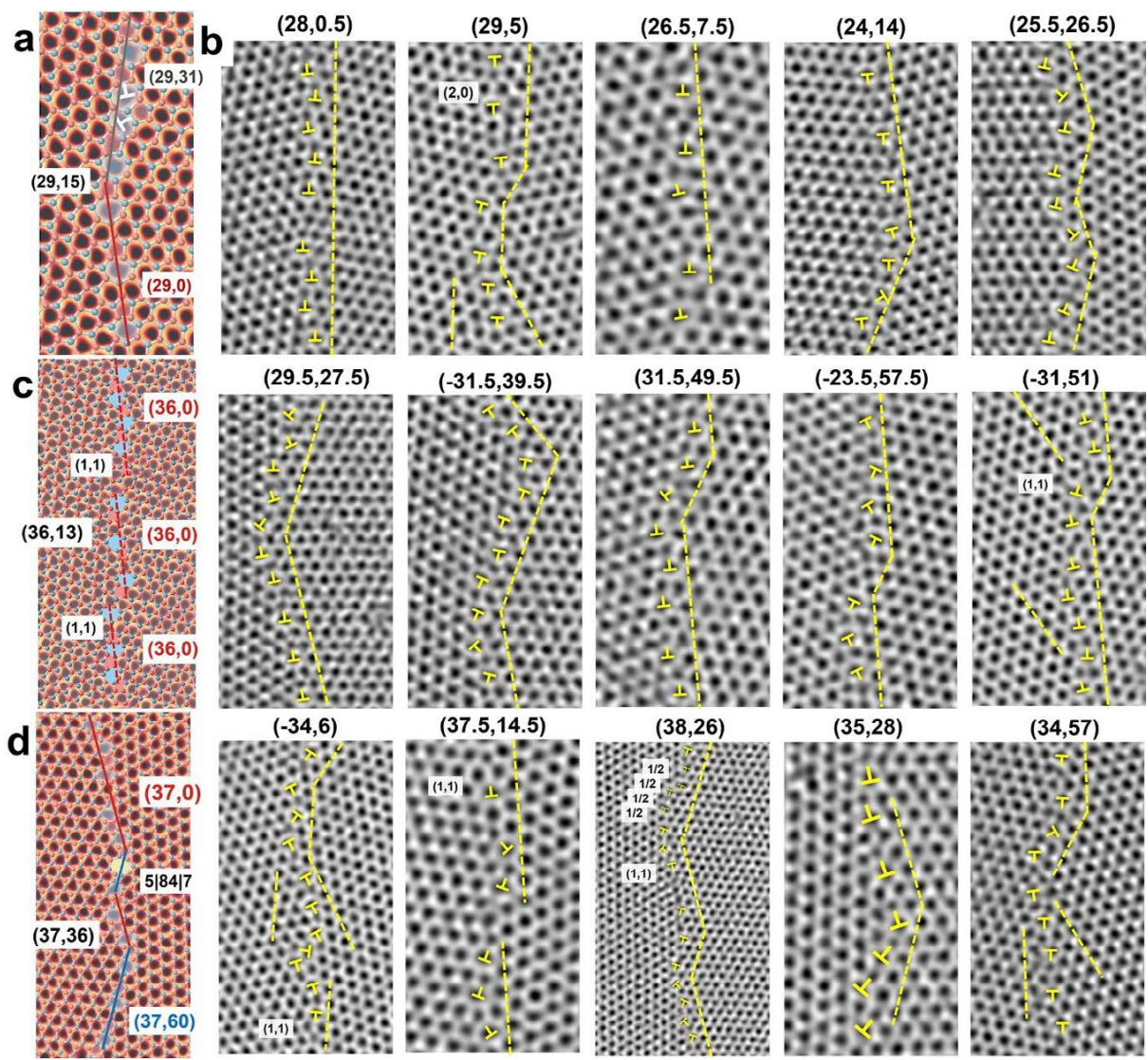

**Figure S17. More examples of GB faceting (|θ|<38°).** The yellow dot lines indicate the symmetric facets.

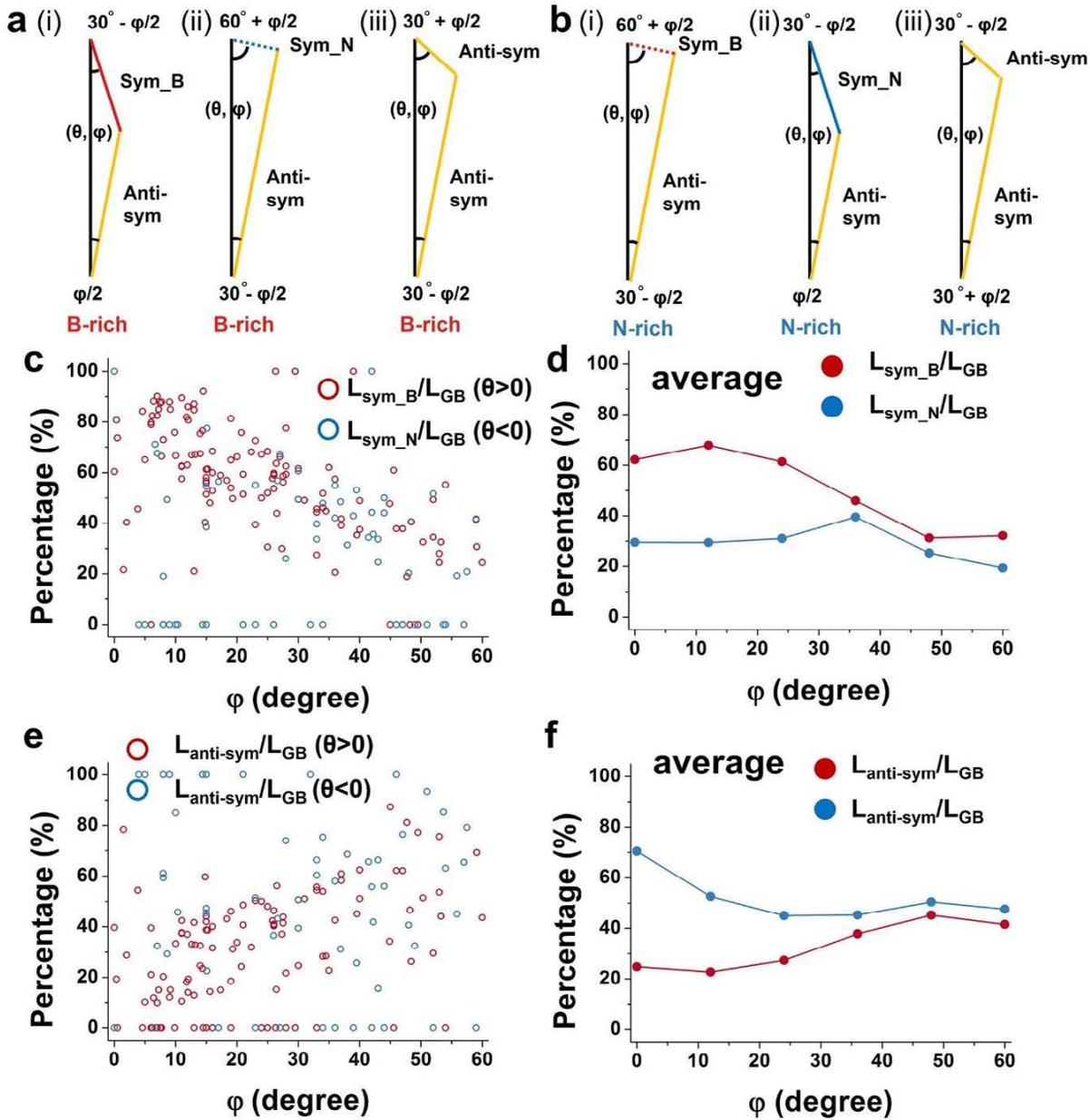

**Figure S18. Differences between faceting in B-rich GBs (θ>0°) and N-rich GBs (θ<0°) (|θ|<38°). a, b,** Schematic diagram of three possible GB faceting in B-rich GBs (**a**) and N-rich GBs (**b**), respectively; (i) for anti-sym facet & symmetric B-rich facet; (ii) for anti-sym facet & symmetric N-rich facet; (iii) for two anti-sym facets, respectively. **c,** The length percentage of symmetric B-rich facets (symmetric N-rich facets) to the overall GB length in θ>0° GBs (θ<0° GB). **e,** The length percentage of anti-symmetric facets to the overall GB length in a θ>0° GB (red) and θ<0° GB (blue), respectively. **d, f,** The average percentage (with a φ range, e.g, 6°<φ<18°) based on the data in **c** and **e**.

## 10.3 More details about GB faceting in |θ|>38° GBs

**Figure S19. GB faceting in GB with |θ|>38°. a-c,** Schematic diagram of three possible GB faceting: a for anti-sym facet & symmetric B-rich facet; **b** for anti-sym facet&symmetric N-rich facet; **c** for two anti-sym facets; (i) for B-rich (θ>0°) and (ii) for N-rich (θ<0°), respectively. **d,** The ratio of overall length of GB to the length of linear GB in three faceting types in **a-c**, respectively. **e-g,** Statistics of GB faceting types as a function of φ. **e** for 38°<|θ|<47°, **f** for 47°<|θ|<60°, and **g** for |θ|=60°, respectively. The symbols

such as "sym_B" represent the inclinations of each facet, and the symbols such as "48&48" represent the configurations that constitute facets.

Three types of faceting in B-rich (θ<0°) and N-rich (θ>0°) condition are displayed in Figure S19a-c, respectively, and their corresponding ratios of the length of overall facets to the straight GB are shown in Figure S19d. Based on this, we ascertained that the faceting types with the smallest ratio in B-rich and N-rich condition are anti-sym facets& symmetric B-rich facets and anti-sym facets& symmetric N-rich facets, respectively. For B-rich (θ>0°) condition, the major faceting is anti-sym facets& symmetric facets (Figure S19e-g), which meets well with the request for the shortest faceting length. However, for N-rich (θ<0°) condition, the major faceting changes to the anti-sym facets& anti-sym facets which are actually not the shortest faceting (Figure S19e-g). Such an asymmetrical behavior in GB faceting of GBs between the θ>0° (B-rich) and θ<0° (N-rich) infers the formation energy of N-N bond should be larger than that of B-B bonds, which is consistent with the results in Figure S18.

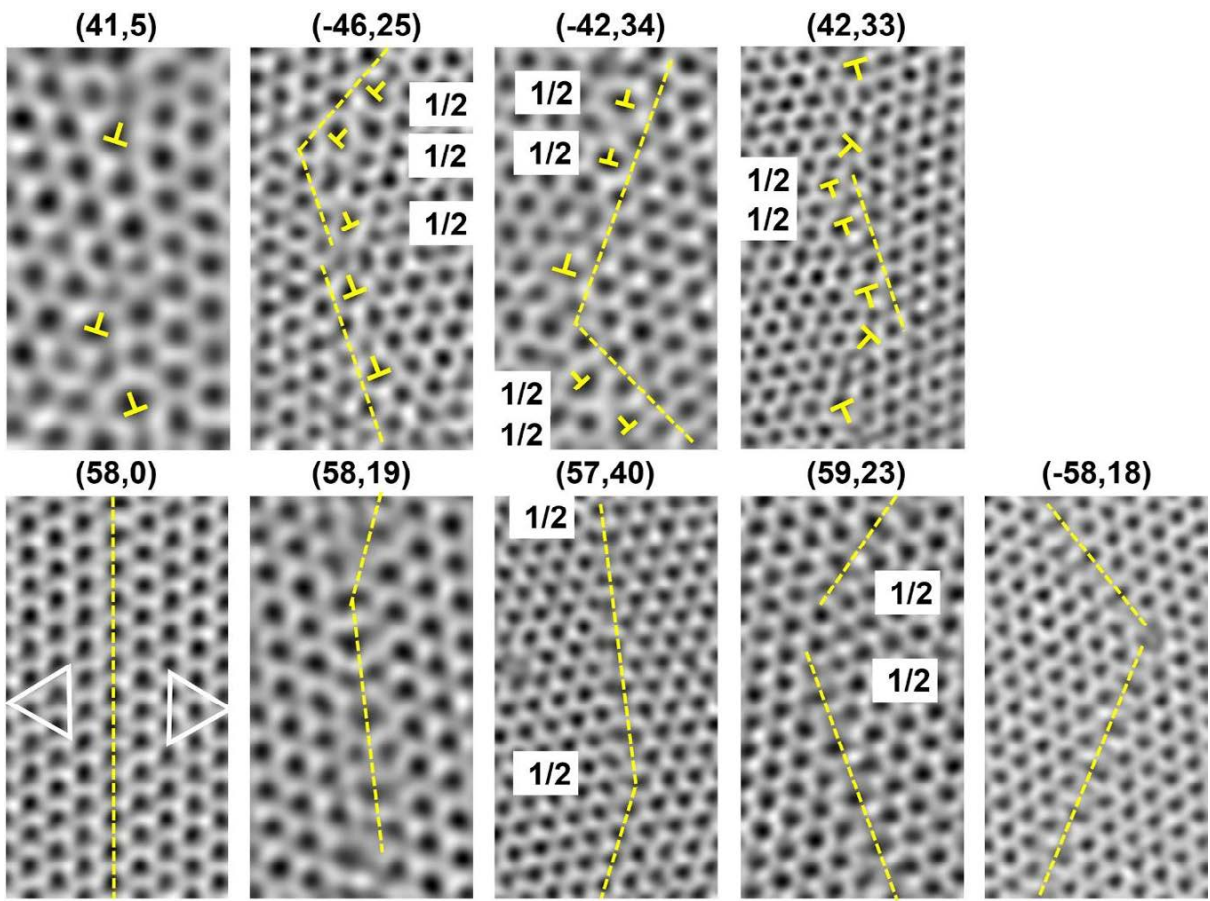

**Figure S20. More examples of GB faceting (38°<|θ|<60°).** The yellow dot lines indicate the symmetric facets.

## 11. The influence of temperature and electron beam on GB structures

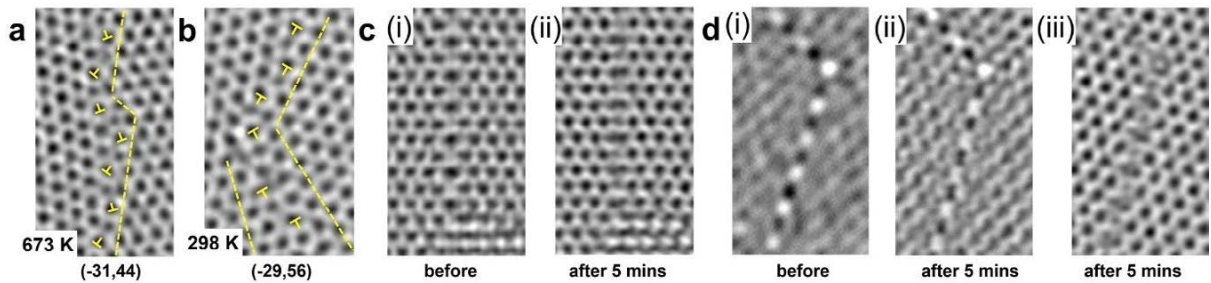

**Figure S21. The effects of temperature and electron on GB structures. a, b,** GB faceting in asym-GBs at lower temperatures (673 K and 298 K). **c, d,** GB-48s and GB-(-60,30) (with two 48-facets) captured before and after the annealing in the absence of electron beam for 5 mins (1073 K). The GB structures remained unchanged during 5-mins annealing process. Note that the different contrasts in d (i), (ii) were due to the deviation of defocus from Scherzer foucs in (iii). It can be seen that the structures remained unchanged by comparing (i) and (ii) (image in (iii) was taken right after (ii)).

## 12. Comparison between the GBs in 2D h-BN and monolayer transition metal dichalcogenides (TMDs)

Both monolayer h-BN and TMDs possess honeycomb lattice, but differently, metal atoms in 2D TMDs are six-fold coordinated and chalcogen atoms are three-fold coordinated, respectively, which actually leads a 3D character in 2D TMD materials. As a result, changing the coordination number of the constituent atoms enriches the structural variety of dislocations in 2D TMDs, compared to that for graphene and 2D h-BN; see Figure S22b. The formation energy of GBs in 2D TMDs have a similar θ dependence with that in 2D h-BN, while it shows a multivalued plots due to the richer variety of stoichiometry of dislocations in 2D TMDs as reported previously[18].

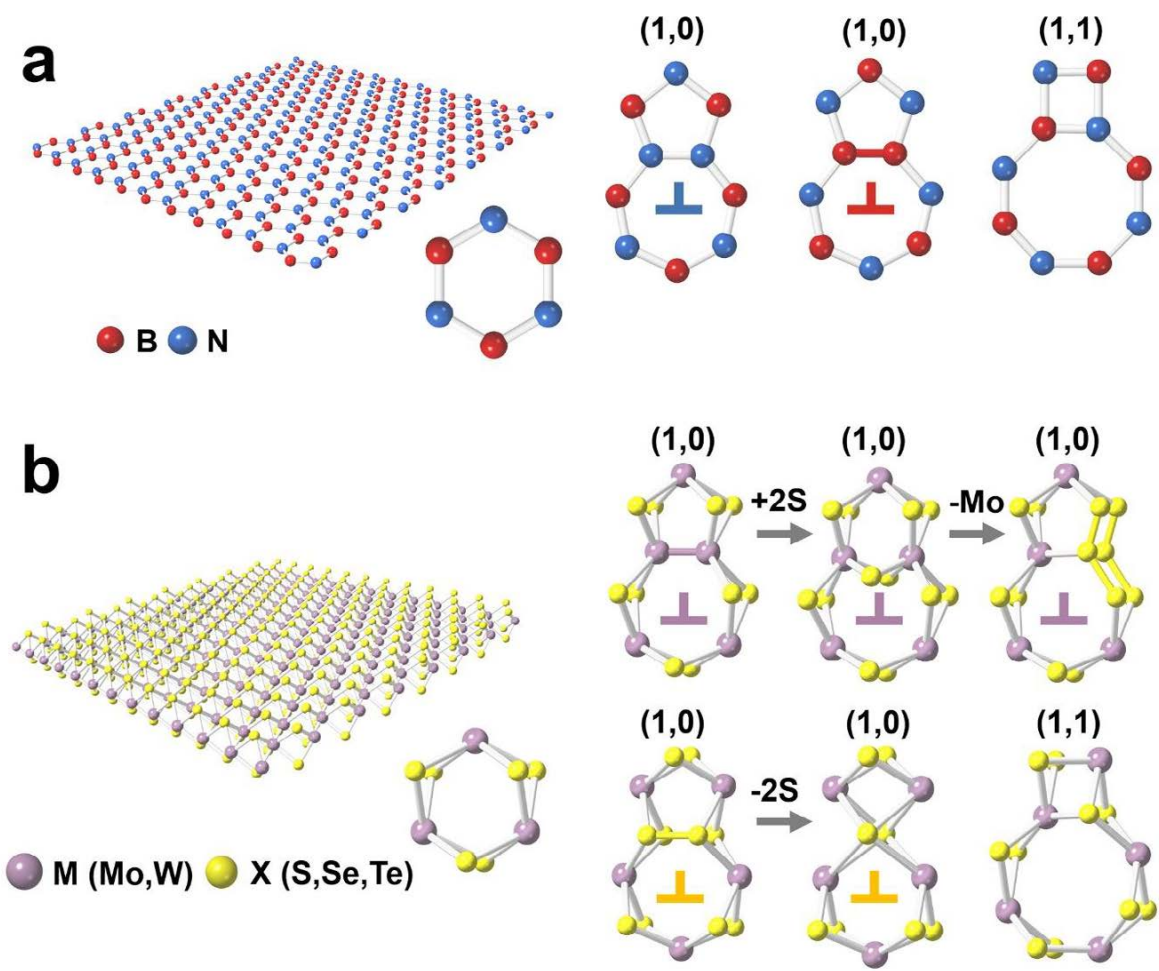

**Figure S22. Dislocations in monolayer h-BN, and monolayer transition metal dichalcogenides (TMDs). a, b,** Structural (left) and dislocation (right) models of monolayer h-BN and TMDs. (1,0) and (1,1) represent the Burgers vectors. Derivative dislocation cores in B were formed by adding or removing metal or chalcogen from the generic dislocation.

  In general, knowledge of GBs in 2D h-BN gained in this study should also be instructive to interpret the GBs in 2D TMDs due to following facts. Firstly, the equilibrium GB structures in 2D TMDs are also a result of balance between the strain and chemical energy of the GBs. From what we learnt from the GBs in 2D h-BN, a critical misorientation angle $\theta_c$ should also exist for 2D TMDs. This angle $|\theta c|=38°$ in 2D h-BN. Similarly in 2D TMDs, for GBs with $|\theta|$ smaller than $\theta_c$, their structures are mainly determined by the interactions of strain fields of the arrays of (1,0) and (0,1) dislocation inside the GB. Secondly, GB faceting should heavily influence the configurations in asym-GBs and sym-GBs (or anti-sym GBs) with an arbitrary θ, which could be partly inferred from the reported results[19,20], but further studies are still needed to resolve the uncertainties.